\documentclass[aps,10pt,prd,showpacs,amsmath,amssymb,twocolumn,preprintnumbers,superscriptaddress,nofootinbib,showkeys]{revtex4-1}

\usepackage[german,english]{babel}
\usepackage{hyperref}
\usepackage{graphicx}
\usepackage{dcolumn}
\usepackage{bm,bbm}
\usepackage{slashed}
\usepackage{MnSymbol}
\usepackage{color}
\usepackage{soul}
\usepackage{lineno}
\usepackage[bottom]{footmisc}

\def\beq{\begin{equation}}
\def\eeq{\end{equation}}
\def\bea{\begin{eqnarray}}
\def\eea{\end{eqnarray}}
\def\nn{\nonumber}

\def\x{{\bm x}}
\def\y{{\bm y}}
\def\x{{\bm x}}
\def\z{{\bm z}}
\def\k{{\bm k}}

\def\p{{\bm p}}
\def\A{{\bm A}}
\def\E{{\bm E}}

\def\D{{\bm D}}

\def\r{{\bm r}}
\def\R{{\bm R}}

\def\0{{\bm 0}}
\def\L{{\bm L}}

\def\bnabla{{\bm \nabla}}
\def\bsigma{{\bm \sigma}}

\def\MS{\overline{\rm MS}}
\def\als{\alpha_{\rm s}}

\begin{document}

\title{The Born-Oppenheimer approximation in an effective field theory language}

\author{Nora Brambilla}
\email{nora.brambilla@ph.tum.de}
\affiliation{Physik-Department, Technische Universit\"at M\"unchen, \\  James-Franck-Str. 1, 85748 Garching, Germany}
\affiliation{Institute for Advanced Study, Technische Universit\"at M\"unchen, \\ Lichtenbergstrasse 2a, 85748 Garching, Germany}
\author{Gast\~ao Krein}
\email{gkrein@ift.unesp.br }
\affiliation{Instituto de F\'{\i}sica Te\'orica, Universidade Estadual Paulista, \\ Rua Dr. Bento Teobaldo Ferraz, 271 - Bloco II, 01140-070 S\~ao Paulo, SP, Brazil}
\author{Jaume Tarr\'us Castell\`a}
\email{jaume.tarrus@tum.de}
\author{Antonio Vairo}
\email{antonio.vairo@ph.tum.de}
\affiliation{Physik-Department, Technische Universit\"at M\"unchen, \\ James-Franck-Str. 1, 85748 Garching, Germany}


\date{\today}

\preprint{TUM-EFT 69/15}

\begin{abstract}
The Born--Oppenheimer approximation is the standard tool for the study of molecular systems. 
It is founded on the observation that the energy scale of the electron dynamics in a molecule is larger than that of the nuclei. 
A very similar physical picture can be used to describe QCD states containing heavy quarks as well as light-quarks or gluonic excitations. 
In this work, we derive the Born--Oppenheimer approximation for QED molecular systems in an effective field theory framework  
by sequentially integrating out degrees of freedom living at energies above the typical energy scale where the dynamics of the heavy degrees of freedom occurs. 
In particular, we compute the matching coefficients of the effective field theory 
for the case of the $H^+_2$ diatomic molecule that are relevant to compute its spectrum up to ${\cal O}(m\alpha^5)$.
Ultrasoft photon loops contribute at this order, being ultimately responsible for the molecular Lamb shift.
In the effective field theory the scaling of all the operators is homogeneous, which facilitates the determination of all the relevant contributions, 
an observation that may become useful for high-precision calculations. 
Using the above case as a guidance, we construct under some conditions an effective field theory for QCD states formed by a color-octet heavy quark-antiquark pair 
bound with a color-octet light-quark pair or excited gluonic state, highlighting the similarities and differences between the QED and QCD systems. 
Assuming that the multipole expansion is applicable, we construct the heavy-quark potential up to next-to-leading order in the multipole expansion 
in terms of nonperturbative matching coefficients to be obtained from lattice QCD.
\end{abstract}

\maketitle

\section{Introduction and Motivation}
\label{sec:intro}
The discovery in the last decade of the $XYZ$ mesons has brought into QCD challenges enduring since the early days of molecular physics in QED 
\textemdash for a recent overview, see Ref.~\cite{Brambilla:2014jmp}. 
A great variety of possible models have been introduced to explain the observed pattern of new mesons. 
A recent proposal~\cite{Brambilla:2008zz,Braaten:2014qka} (see also~\cite{Vairo:2009tn}) advocates the use of the
Born--Oppenheimer (BO) approximation~\cite{bo1,bo2,bo3,RevModPhys.35.473}, familiar to QED molecular physics, 
as a starting point for a coherent description of the new QCD structures. 
The rational for this being that many of the new mesons contain a heavy quark-antiquark pair, 
and the time scale for the evolution of the gluon and light-quark fields is small compared to that for the motion of the heavy quarks. 
Although the BO approximation has been used in the past to study heavy hybrids by means of quenched lattice data for gluonic static potentials~\cite{Juge:1997nc,Griffiths:1983ah,Meyer:2015eta}\footnote{Models have also been used for the determinations of the gluonic static potentials and heavy hybrids in a BO framework, see for example Refs.~\cite{Guo:2007sm,Guo:2008yz}}, 
the new aspect of the proposal in Refs.~\cite{Brambilla:2008zz,Braaten:2014qka} is the recognition that
the BO approximation can also be applied to mesons with light quark and antiquark flavors when input from lattice simulations becomes available. 
In the present paper we go one step further in this proposal and develop an effective field theory (EFT) 
that allows to calculate in a systematic and controlled manner corrections to the BO approximation for QED and QCD molecular systems.

An EFT is built by sequentially integrating out degrees of freedom induced by energy scales higher than the energy scale we are interested in. 
For QED molecules, such a sequential process proceeds as follows: 
(A) integrating out {\em hard} modes associated with the masses of the charged particles leading to nonrelativistic QED (NRQED)~\cite{Caswell:1985ui,Kinoshita:1995mt}, 
(B) integrating out {\em soft} modes associated with the relative momenta between electrons and nuclei in NRQED leading to potential NRQED (pNRQED)~\cite{Pineda:1997bj,Pineda:1997ie},
and (C) exploiting the fact that the nuclei move much slower than the electrons due to their heavier masses, 
modes associated with the electron and photon dynamics at the electron binding energy scale, the {\em ultrasoft} scale, 
can be integrated out leading to an EFT for the motion of the nuclei only. 
In QED these steps can be done in perturbation theory.

In the present paper we compute this ultimate EFT in the simple case of a QED molecule formed by two heavy nuclei and one electron, like the $H^+_2$ ion molecule. 
Because the BO approximation emerges as the leading-order approximation in this EFT, we call it Born--Oppenheimer EFT (BOEFT).
Furthermore we show how the EFT allows to systematically improve on the leading-order approximation 
by calculating corrections in the inverse of the mass of the nuclei as well as electromagnetic corrections.
We give explicit analytical expressions, regularized in dimensional regularization when needed,  
for the different contributions to the binding energy of the two nuclei plus one electron molecule up to ${\cal O}(m\alpha^5)$.
It is at this order that the Lamb shift is generated. 

The BOEFT that we construct is new, although NRQED has been applied in atomic and molecular physics for nearly two decades~\cite{Kinoshita:1995mt,Labelle:1996en}. 
In particular, NRQED has been used for computing the leading relativistic, recoil and radiative corrections to the energy levels of the $H^+_2$ molecule in Ref.~\cite{PhysRevA.74.052506} 
and for computing higher-order corrections in Refs.~\cite{Jentschura:2005xu,PhysRevA.74.040502,PhysRevA.77.022509,PhysRevA.79.012501,PhysRevA.87.062506}. 
The new and distinctive aspect of our approach is that we carry out the full EFT program for the diatomic molecule, 
integrating out not only the hard scale, as in NRQED, but also the soft and ultrasoft scales. 
The advantage is that each term in the Lagrangian has a unique size and the scaling of Feynman diagrams is homogeneous. 
This greatly facilitates the determination of all the relevant contributions to a given observable up to a given precision, 
a feature that is particularly useful for higher-order calculations.

An analog EFT for QCD states containing a heavy quark-antiquark pair in a color-octet state bound with light quarks or a gluonic color-octet state can be built following a similar path. 
However, unlike QED molecules, the QCD states are determined by nonperturbative interactions. 
The hard scale set by the heavy-quark mass can always be integrated out perturbatively, leading to nonrelativistic QCD (NRQCD)~\cite{Caswell:1985ui,Bodwin:1994jh}. 
At short enough distances the relative momentum of the heavy quarks can also be integrated out perturbatively 
resulting in potential nonrelativistic QCD (pNRQCD)~\cite{Pineda:1997bj,Brambilla:1999qa,Brambilla:1999xf,Brambilla:2004jw}.\footnote{
A strongly-coupled version of pNRQCD for the case when the soft scale is nonperturbative 
has been worked out in~\cite{Brambilla:2000gk,Pineda:2000sz,Brambilla:2003mu}
In this case, the matching coefficients are written in terms of gauge invariant Wilson loops with field insertions.}. 
Similarly to the diatomic molecule case, the heavy quarks move slower than the light degrees of freedom, 
whose spectrum is assumed to appear at the scale $\Lambda_{\rm QCD}$. 
Thus, one can construct an EFT for these QCD ``molecular'' states by integrating out the scale $\Lambda_{\rm QCD}$. 
Since this is the scale of nonperturbative physics, the matching coefficients will be nonperturbative quantities to be determined, for instance, by lattice calculations. 
When light quarks are neglected, one regains in this way the EFT recently constructed for quarkonium hybrids~\cite{Berwein:2015vca}.

The paper is organized as follows. 
In Sec.~\ref{sec:pNRQED} we construct the pNRQED Lagrangian for two nuclei and one electron. 
In Sec.~\ref{sec:HHL} we proceed with integrating out the ultrasoft scale and constructing the molecular EFT, BOEFT. 
Section~\ref{sec:self} is devoted to the power counting of the BOEFT, 
which we use to assess the importance of the nonadiabatic coupling and other corrections to the molecular energy levels. 
The EFT for the QCD analog of the diatomic molecule, quarkonium hybrids and tetraquark mesons built out of a heavy quark and antiquark, is developed in Sec.~\ref{sec:BO-QCD}. 
Section~\ref{sec:concl} contains the conclusions and an outlook for future developments. 
The Appendix presents a detailed calculation of the Lamb shift for the $H^+_2$ molecule.

\section{\texorpdfstring{\lowercase{p}}{p}NRQED}
\label{sec:pNRQED}
We aim at building an EFT for a molecular system containing heavy and light particles:
the heavy particles (nuclei) have electric charge $+Ze$ and mass $M$ 
and the light particles (electrons) have electric charge $-e$ and mass $m$, with $M~\gg~m$. 
Both kinds of particles are nonrelativistic. 
Such a molecular system has several well-separated energy scales, as we will see more in detail in the following. 
From the highest to the lowest one the relevant energy scales are 
the masses of the heavy and light constituents (hard scales), 
the typical relative momentum $p = |{\p}| \sim m v$ between heavy and light particles (soft scale) 
and the binding energy of the light particles $E \sim m v^2$ (ultrasoft scale). 
For a Coulomb-type interaction it holds that $v\sim \alpha$ with $\alpha = e^2/4\pi \sim 1/137$ the fine structure constant. 
Finally, specific of molecules an extra low-energy scale appears: the binding energy of the heavy nuclei.

The EFT suitable for describing QED bound states at the ultrasoft scale is pNRQED. 
In Ref.~\cite{Pineda:1997ie} it was worked out for the hydrogen atom, in this section we extend pNRQED to describe systems with two nuclei and one electron. 
In Sec.~\ref{sec:HHL} we will integrate out the ultrasoft modes and build the EFT suitable to describe the molecular states.

The Lagrangian of pNRQED can be written in terms of the light and heavy fermion fields, $\psi(t,\x)$ and $N(t,\x)$ respectively, and the ultrasoft-photon field, $A_\mu(t,\x)$. 
The meaning of $A_\mu(t,\x)$ being ultrasoft is that it must be multipole expanded (e.g., about the position of the center of mass (c.m.) of the constituents). 
The operators of the pNRQED Lagrangian can be organized in an expansion in $\alpha$ and $m/M$. 
In order to homogenize the counting in these two expansion parameters, we will use that $m/M$ is numerically similar to $\sim\alpha^{3/2}$. 
Then, the pNRQED Lagrangian relevant to compute the spectrum up to order ${\cal O}(m\alpha^5)$ reads
\begin{widetext}
\bea
L_{\rm pNRQED} &=& - {\frac 1 4} \int d^3 x \, F_{\mu \nu}(x) F^{\mu \nu}(x) 
+  \int d^3 x \, \bigl\{\psi^{\dagger}(t,\x) \left[i \, \partial_t - h_{e}(t,\x) \right] \psi(t,\x) 
+ N^{\dagger}(t,\x) \left[ i \, \partial_t  - h_{Ze}(t,\x) \right]N(t,\x) \bigr\} 
\nn \\
&& -\, \int d^3 x \, d^3y \, \Bigl[ \psi^\dagger(t,\x) \psi (t,\x)  V_{Ze}(\x - \y, {\bm \sigma}) +  N^{\dagger}(t,\x) N(t,\x) V_{ZZ}(\x - \y) \Bigr] N^\dagger(t,\y)  N(t,\y)\,,
\label{LNpsi}
\eea
\end{widetext}
where $F_{\mu\nu} = \partial_\mu A_\nu - \partial_\nu A_\mu$ and all photons are ultrasoft. 
Moreover we have used 
\bea
&& \hspace{-0.5cm}h_{e}(t,\x) = eA_0(t,\x) - \frac{{\D}^2_{-e}(t,\x)}{2 m} - \frac{{\D}^4_{-e}(t,\x)}{8 m^3}\,, 
\label{he} \\
&& \hspace{-0.5cm}h_{Ze}(t,\x) = - Ze A_0(t,\x) - \, \frac{{\D}^2_{Ze}(t,\x)}{2 M}\,,
\label{hZ}
\eea
where ${\D}_q$ is the covariant derivative, with $q = - e$ for the electron and $q = + Ze$ for the nuclei:
\bea
i{\D}_q(t,\x) = i{\bm \nabla}_x - q {\A}(t,\x)\,.
\label{Dq}
\eea
The electron-nucleus potential $V_{Ze}(\x, \sigma)$ is given by
\beq
V_{Ze}(\x,\sigma) = V^{\rm LO}_{Ze}(\x) +  V^{\rm NLO}_{Ze}(\x, \sigma_e)\,,
\label{Ve}
\eeq
where ${\rm LO}$ (leading order) and ${\rm NLO}$ (next-to-leading order) refer to the order $m\alpha^2$ and $m\alpha^4$ contributions to the spectrum respectively.
The LO potential is the Coulomb potential
\beq
V^{\rm LO}_{Ze}(\x)  = -  \frac{Z\alpha}{|\x|}\,, 
\label{Ve-LO}
\eeq
while the NLO one is the sum of a contact and spin-orbit interaction
\beq
V^{\rm NLO}_{Ze}(\x, \sigma_e) = V^{\rm ct}_{Ze}(\x) + V^{\rm SO}_{Ze}(\x, \sigma_e)\,,
\label{Ve-NLO}
\eeq
with
\bea
V^{\rm ct}_{Ze}(\x) &=&  -\frac{Z \alpha}{m^2} \left(-\frac{c_D}{8} + 4 d_2 \right) \, 4 \pi \delta(\x)\,, 
\label{VCT} \\
V^{\rm SO}_{Ze}(\x,\sigma_e) &=& - i c_S \frac{Z\alpha}{4m^2} \, {\bsigma}_e 
\cdot  \left(\frac{\x}{|\x|^3} {\bf \times} \bnabla_x \right)\,,
\label{VSO}
\eea
where $c_D$, $c_S$ and $d_2$ are matching coefficients that up to order $\alpha$ read 
\beq
c_D = 1 + \frac{\alpha}{\pi} \left(\frac{8}{3} \log \frac {m}{\mu}\right),
\hspace{0.5cm} c_S = 1 + \frac{\alpha}{\pi}, \hspace{0.5cm} d_2 = \frac{\alpha}{60\pi}\,.
\label{cdcsd2}
\eeq
The coefficient $c_D$ has been renormalized in the $\MS$ scheme.
The scale $\mu$ is the dimensional regularization scale that in the case of $c_D$ acts as an infrared factorization scale.
Finally, the $V_{ZZ}$ potential in Eq.~(\ref{LNpsi}) contains the LO nucleus-nucleus Coulomb potential:
\beq
V_{ZZ}(\x)  = \frac{Z^2\alpha}{|\x|}\,.
\label{VZ-LO}
\eeq
Further contributions to \eqref{Ve} and \eqref{VZ-LO}, which can be found in Ref.~\cite{Peset:2015zga}, are beyond our accuracy.

Next, we project the Lagrangian in Eq.~(\ref{LNpsi}) on the subspace of one electron and two nuclei.
This is similar to the pNRQED bound state calculations for the hydrogen atom~\cite{Pineda:1997bj,Pineda:1997ie},
but since the projection for one light and two heavy particles with different charges has not been done so far in the literature,
we present the procedure with some detail.
The subspace of one electron and two nuclei is spanned by Fock-space states of the form
\bea
|\varphi(t) \rangle &=& \int d^3x d^3y_1 d^3y_2 \, \varphi(t,\x,\y_1,\y_2) \nn \\
&& \times \, \psi^\dag(t,\x) N^\dag(t,\y_1) N^\dag(t,\y_2)|{\rm US}\rangle\,,
\label{state12}
\eea
where $\varphi(t,\x,\y_1,\y_2)$ is the wave function of the system 
and $|{\rm US}\rangle$ is the Fock-space state containing no hard particles (electrons or nuclei) and an arbitrary number of ultrasoft ones (photons).
The corresponding projected Lagrangian, adequate for calculating the spectrum up to ${\cal O}(m\alpha^5)$, is
\begin{widetext}
\bea
L_{\rm pNRQED}&=&-{\frac 1 4}\int d^3 x \,F_{\mu \nu}(x) F^{\mu \nu}(x)+\int d^3x \, d^3y_1 \, d^3y_2 \, 
\varphi^\dag(t,\x,\y_1,\y_2)\bigl[i\,\partial_t- h_{e}(t,\x)-h_{Ze}(t,\y_1)-h_{Ze}(t,\y_2) \nn \\
&& - \, V_{Ze}(\x - \y_1,\bsigma)  -  V_{Ze}(\x - \y_2,\bsigma) - V_{ZZ}(\y_1 - \y_2) \bigr]\varphi(t,\x,\y_1,\y_2)\,,
\label{LNNpsi}
\eea
\end{widetext}
where we have promoted $\varphi(t,\x,\y_1,\y_2)$ to a tri-local field.

To ensure that the photon fields $A_\mu$ are ultrasoft one may multipole expand them about the c.m. of the system. 
The task is facilitated by defining an appropriate c.m. and relative coordinates.
The c.m. coordinate $\R$ of the system is given by
\beq
\R = \frac{m \x + M (\y_1 + \y_2)}{m + 2M}\,.
\label{R3p}
\eeq
To describe the motion of the electron relative to the positions $\y_1$ and $\y_2$ of the nuclei we use
\bea
{\bm \z} = \x - \frac{\y_1+\y_2}{2}\,,
\label{relat}
\eea
and for the relative coordinate of the nuclei 
\beq
{\bm r} = \y_1 - \y_2\,.
\label{r}
\eeq

The multipole expansion spoils manifest gauge invariance.
It is important, however to recall that we have an EFT for ultrasoft gauge fields, hence gauge transformations must not introduce into the EFT
gauge fields with large-momentum components; that is, the allowed gauge transformations are those that produce fields that still are within the EFT.
One can recover manifest (ultrasoft) gauge invariance at least for charge neutral systems by introducing the field redefinition:
\bea
\varphi(\x,\y_1,\y_2,t) &=& U_{-e}(\x,\R,t) U_{+Ze}(\y_1,\R,t) U_{+Ze}(\y_2,\R,t)
\nn \\
&& \times \, S(t,\x,\y_1,\y_2,t)\,, 
\label{defSHHl}
\eea
where $U_q$ is the Wilson line
\bea
U_{q}(\x,\R,t) = e^{i q\int^{\R}_{\x} d\x' \cdot \A(\x',t)}\,.
\label{defU}
\eea
Under a gauge transformation $A_0(t,\R) \rightarrow A_0(t,\R)- \partial_t \theta(t,\R)$
and $\A(t,\R) \rightarrow \A(t,\R)+{\bm \nabla}_R \theta(t,\R)$, the field $S(t,\R,\r,\z)$ transforms as 
\beq
S(t,\R,\r,z) \rightarrow e^{-i e_{\rm tot} \theta(t,\R)} \, S(t,\R,\r,\z)\,,
\label{gauge-transf}
\eeq
where $e_{\rm tot}$ is the total charge:
\beq
e_{\rm tot} = -e \, (1 - 2Z)\,,
\label{e_r} 
\eeq
For a charge-neutral system, $e_{\rm tot} = 0$, and the field $S(t,\R,\r,z)$ is gauge invariant.

The Lagrangian in terms of the field $S$ is given by
\begin{widetext}
\bea
L_{\rm pNRQED}  &=& - {\frac 1 4} \int d^3 x \, F_{\mu \nu}(x) F^{\mu \nu}(x) + \int d^3R \, d^3r \, d^3z \, S^\dag(t,\R,{\r},{\z})
\biggl[i \, \partial_t +  \frac{{\bm \nabla}^2_{\! R} }{2{M_{\rm tot}}} + \,  e_{\rm tot} A_0(t,\R) + e_{\rm eff} \, {\z} \cdot {\E}(t,\R)
  \nn \\
&&- \, h_0(\r,\z) + \frac{{\bm \nabla}^2_{r}}{M}  - V^{\rm LO}_{ZZ}(\r) 
  + \frac{{\bm \nabla}^2_z }{4M}+\frac{{\bm \nabla}^4_z}{8m^3} - \, V^{\rm NLO}_{Ze}(\z+\r/2,\bsigma)-\,V^{\rm NLO}_{Ze}(\z-\r/2,\bsigma) \biggr]S(t,\R,{\r},{\z})\,,
\label{LNpsigi-HHl}
\eea
\end{widetext}
where 
\bea
h_0(\r,\z) &=& -  \frac{{\bm \nabla}^2_z }{2m} + V^{\rm LO}_{Ze}(\z+\r/2)+ V^{\rm LO}_{Ze}(\z-\r/2)\,,\label{h0}
\eea
with ${M_{\rm tot}}$ being the total mass
\bea
{M_{\rm tot}} = m + 2M \,,
\label{redmass-lHH}
\eea
${\E}(t,\R) = - \partial_t \A(t,\R) - {\bm \nabla}_R A_0(t,\R)$ is the electric field and $e_{\rm eff}$ is the effective charge:  
\beq
e_{\rm eff} = 2 e \, \frac{M + Zm}{m + 2M}= e + {\cal O}(\alpha^2)\,.
\label{e_eff}
\eeq

The sizes of the different terms that appear in the Lagrangian \eqref{LNpsigi-HHl} are as follows.
\begin{enumerate}
\item Relative electron-nuclei momentum $-i{\bm \nabla}_z$ and inverse relative distance $1/|\z|$ have size $m \alpha$.
\item Photon fields, derivatives acting on photon fields, the time derivative, and c.m. momentum, $-i{\bm \nabla}\!_R$, acting on $S$ have size $m\alpha^2$.
\item As we shall discuss in Sec.~\ref{sec:self}, the inverse relative nuclei-nuclei distance is $1/r \sim m\alpha$,
  whereas the radial part of the derivative ${\bm \nabla}_r \sim (M/m)^{1/4} \, m \alpha\sim m\alpha^{5/8} $ when acting on the nuclei,
  but ${\bm \nabla}_r \sim m\alpha$ when acting on the electron cloud. This implies that the kinetic energy associated with the
  relative motion of the nuclei is $-\bm{\nabla}^2_{r}/M \sim m \alpha^2 \sqrt{m/M} \sim m\alpha^{11/4}$.
\end{enumerate}

Using this counting, and disregarding operators that produce emission or absorption of photons that contribute only in loops,
the leading-order operators in Eq.~(\ref{LNpsigi-HHl}) are $h_0(\r,\z)+V^{\rm LO}_{ZZ}(\r)$, which are of $\mathcal{O} \left(m\alpha^2\right)$.
Since the kinetic energy associated with the relative motion of the two nuclei, $-\bm{\nabla}^2_{r}/M$, is of $\mathcal{O}\left(m\alpha^{11/4}\right)$,
at leading order the nuclei are static and $V^{\rm LO}_{ZZ}(\r)$ is just a constant.
Therefore, at leading order, the Euler--Lagrange equation from the Lagrangian \eqref{LNpsigi-HHl} is nothing else than a Schr\"odinger equation
for the electronic energy levels with Hamiltonian $h_0(\r,\z)$.
Corrections to these energy levels can be obtained in perturbation theory.
Parametrically, the first of such corrections is given by the recoil term, ${\bm \nabla}^2_z/4M$, which is $\mathcal{O}\left(m\alpha^{7/2}\right)$,
and the second one by $\bm{\nabla}^4_z/8m^3+V^{\rm NLO}_{Ze}$, which starts at $\mathcal{O}\left(m\alpha^{4}\right)$.
The $\mathcal{O}\left(m\alpha^{5}\right)$ corrections include the Lamb shift, and originate from ultrasoft photon loops
and subleading contributions to the NLO potentials.

To obtain the molecular energy levels we need to solve the dynamics of the $r$ coordinate.
In principle we could do this by adding subleading terms to the Hamiltonian,
$h_0(\r,\z) + \bm{\nabla}^2_{r}/M+V^{\rm LO}_{ZZ}(\r) + \dots$, and solving the corresponding Schr\"odinger equation.
However, in this paper, following the logic of EFTs,
we will integrate out from pNRQED the ultrasoft degrees of freedom to obtain an EFT at the energy scale of the two-nuclei dynamics.
The Euler--Lagrange equation of this EFT provides a Schr\"odinger equation for the molecular energy levels.
We will develop this EFT, which we call BOEFT, in the following section.

Since the c.m. motion does not affect the internal dynamics of the molecule, we can simply work in the c.m. frame and ignore the dependence on $\R$ of the field $S$.
We also use the notation $A_0(t,\0)$ and $\E(t,\0)$ to indicate quantities defined at the origin of the coordinate system, i.e., $\R = \0$.

\section{Born--Oppenheimer EFT for diatomic molecules} 
\label{sec:HHL}
Our purpose is to build the BOEFT, an EFT for the diatomic molecule at the energy scale of the two-nuclei dynamics. 
This EFT is obtained by integrating out the ultrasoft scale, $m\alpha^2$, from pNRQED for two nuclei and one electron given in Sec.~\ref{sec:pNRQED}.
We will include effects that contribute to the binding energy of the molecule up to~${\cal O}(m\alpha^5)$.

Since the electron dynamics occurs at the ultrasoft scale, integrating out this scale entails that all the electronic degrees of freedom are integrated out. 
Moreover, also ultrasoft photons are integrated out. 
Therefore, the degrees of freedom of the BOEFT are nuclei and photons with energies of $\mathcal{O}\left(m\alpha^{11/4}\right)$ or smaller.

The tree-level matching contributions can be easily obtained by expanding the field $S(t,\r,\z)$ in the pNRQED Lagrangian of Eq.~(\ref{LNpsigi-HHl}) 
in eigenfunctions of the leading-order Hamiltonian $h_0(\r,\z)$ of Eq.~(\ref{h0}). This corresponds in expanding the field $S(t,\r,\z)$ as

\beq
S(t,\r,\z) = \sum_{\kappa} \Psi_{\kappa}(t,\r) \, \phi_{\kappa}(\r;\z)\,,
\label{exp-HHl}
\eeq
where $\phi_{\kappa}(\r;\z) = \langle \z| \r,\kappa\rangle$ satisfy the electronic eigenvalue equation 
\beq
h_0(\r,\z) \phi_{\kappa}(\r;\z) = V^{\rm light}_{\kappa}(\r) \, \phi_{\kappa}(\r;\z)\,.
\label{eigen-HHl}
\eeq

The eigenvalues $V^{\rm light}_{\kappa}(\r)$ are the static energies, with~$\kappa$ representing the set of quantum numbers 
specifying the electronic state for a fixed separation~$\r$ of the nuclei. 
The $\r$ in the state vector $|\r,\kappa\rangle$ emphasizes that eigenvalues labeled by $\kappa$ refer to a given nuclei separation~$\r$. 
The eigenfunctions $\phi_{\kappa}(\r;\z)$ are orthonormal:
\beq
\int d^3z \, \phi^*_{\kappa}(\r;\z) \phi_{\kappa'}(\r;\z)= \delta_{\kappa\kappa'}\,.
\label{ortho-HHl}
\eeq
The static electronic energies $V^{\rm light}_{\kappa}(\r)$ scale like $m\alpha^2$.

The set of quantum numbers $\kappa$ is familiar from molecular physics
and corresponds to representations of the symmetry group of a diatomic molecule~\cite{LandauLifshitz}: 
the eigenvalue $\lambda = 0, \pm 1,\cdots $ of the projection of the electron angular momentum on the axis joining the two nuclei, $\hat r$, 
traditionally denoted by $\Lambda = |\lambda|$ and conventionally labeled by $\Sigma,\, \Pi,\, \Delta,\, \dots$ for $\Lambda = 0,\,1,\,2,\,\dots$;
the total electronic spin~$S$, with the number of states (multiplicity) for a given~$S$ being $2S+1$, and indicated with an index, like $^{2S+1}\Sigma$; 
additionally, for the $\Sigma$ state, there is a symmetry under reflection in any plane passing through the axis $\hat r$, 
the eigenvalues of the corresponding symmetry operator being $\pm 1$ and indicated as $\Sigma^{\pm}$; 
and, in the situation of identical heavy nuclei, the eigenvalues $\pm 1$ of the parity operator of reflections through the midpoint between the two nuclei,
denoted by $g = + 1$ and $u = - 1$.\footnote{
In the heavy quark-antiquark case that we will discuss in Sec.~\ref{sec:BO-QCD}, the parity operator is replaced by the CP operator.} 
In this way, a possible ground state is denoted by $\kappa = \, ^1\Sigma^{+}_g$.

The tree-level matching is sufficient up to terms in the Lagrangian of ${\cal O}(m\alpha^4)$.
Ultrasoft photon loops start contributing at ${\cal O}(m\alpha^5)$ and are responsible for the Lamb shift of the diatomic molecule. 
We detail the calculation of the leading ultrasoft loop in Appendix~\ref{app:matching}.

The BOEFT Lagrangian up to ${\cal O}(m\alpha^5)$ reads
\begin{widetext}
\bea
L_{\rm BOEFT} &=& - {\frac 1 4} \int d^3 x \, F_{\mu \nu}(x) F^{\mu \nu}(x) 
\nn \\
&& + \, \int d^3r \, \sum_{\kappa \kappa'} \Psi^\dag_{\kappa}(t,\r) \bigl\{ \bigl[ i \partial_t + e_{\rm tot} A_0(t,\0) 
  - \, H^{(0)}_\kappa(\r) - \delta E_\kappa(\r)\bigr] \delta_{\kappa\kappa'} - C^{\rm nad}_{\kappa\kappa'}(\r)  \bigr\} \Psi_{\kappa'}(t,\r)\,.
\label{Lag-fin}
\eea
The photon fields carry energies and momenta of $\mathcal{O}\left(m\alpha^{11/4}\right)$ or smaller.
The operator $H^{(0)}_\kappa$ is the leading-order nuclei-nuclei Hamiltonian: 
\beq
H^{(0)}_\kappa(\r) =  - \frac{\bnabla^2_r}{M} + V^{\rm LO}_{ZZ}(\r)+ V^{\rm light}_{\kappa}(\r)\,,
\label{H0}
\eeq
and $\delta E_\kappa(\r)$ is the sum of the tree-level and second order recoil, Breit--Pauli corrections as well as the one-loop ultrasoft one:
\beq
\delta E_\kappa(\r) = \delta^{\rm rec} E_\kappa(\r)+\delta^{{\rm rec},\,2} E_\kappa(\r)+\delta^{\rm NLO} E_\kappa(\r)+\delta^{\rm US} E_\kappa(\r)\,.
\label{delta-E}
\eeq
The counting of $H^{(0)}_\kappa$  will be justified in the next section, but we have already anticipated
that the eigenvalues of $H^{(0)}_\kappa$ scale like $m\alpha^2 \sqrt{m/M} \sim m \alpha^{11/4}$.

The different contributions to $\delta E_\kappa(\r)$ read
\bea
\delta^{\rm rec} E_\kappa(\r) &=&  \int d^3z\,\phi^*_\kappa(\r;\z) \,\left(- \frac{{\bm \nabla}^2_z}{4M}\right)\, \phi_\kappa(\r;\z)
= \langle \r, \kappa |(- {{\bm \nabla}^2_z})/({4M})|\r,\kappa\rangle, 
\label{E-rec}
\eea
which is of order $m\alpha^2 \, m/M \sim m \alpha^{7/2}$, 
\bea
\delta^{{\rm rec},\,2} E_\kappa(\r) &=& 
\sum_{\bar\kappa\neq \kappa} \frac{|\langle \r, \kappa |(- {{\bm \nabla}^2_z})/({4M})|\r,\bar\kappa\rangle|^2}{V^{\rm light}_{\kappa}(\r) - V^{\rm light}_{\bar\kappa}(\r)}\,,
\label{E-rec2}
\eea 
which is of order $m\alpha^2 \, (m/M)^2 \sim m \alpha^5$, 
\bea
\delta^{\rm NLO} E_\kappa(\r)  &=& \int d^3z \, \phi^*_{\kappa}(\r;\z)\biggl[V^{\rm NLO}_{Ze}(\z+\r/2, \sigma) 
+ \, V^{\rm NLO}_{Ze}(\z-\r/2, \sigma)- \frac{\nabla^4_z}{8m^3}\biggr]\phi_{\kappa}(\r;\z) ,
\label{E-NLO} 
\eea
which starts at order $m\alpha^4$, and 
\bea
\delta^{\rm US} E_\kappa(\r) &=& - \frac{e^2}{6 \pi^2} \biggl\{ - \frac{Z e^2}{2m^2}\, 
\left[ \log \left(\frac{\mu}{m}\right) + \frac{5}{6} - \log (2)\right]\, \rho_\kappa(\r)
\nn \\
&& + \sum_{\bar\kappa\neq \kappa} |\langle \r, \kappa |{\bm v}_z|\r,\bar\kappa\rangle|^2 \, 
(V^{\rm light}_{\kappa}(\r) - V^{\rm light}_{\bar\kappa}(\r)) \, \log\left(\frac{m}{|V^{\rm light}_{\kappa}(\r) 
- V^{\rm light}_{\bar\kappa}(\r)|}\right)\biggr\}\,,
\label{Eultrasoft} 
\eea
where ${\bm v}_z = -i\,{\bm \nabla}_z/m$, $\langle \r,\kappa|{\bm v}_z|\r,\bar\kappa\rangle$ is the matrix element
\beq
\langle \r,\kappa|{\bm v}_z|\r,\bar\kappa\rangle = \int d^3z \, \phi^*_{\kappa}(\r;\z) \,{\bm v}_z\, 
\phi_{\bar\kappa}(\r;\z)\,,
\eeq
and $\rho_\kappa(\r)$ is the electron density at the positions of the nuclei
\beq
\rho_\kappa(\r) = |\phi_\kappa(\r;\z=\r/2)|^2+|\phi_\kappa(\r;\z=-\r/2)|^2\,.
\eeq
The ultrasoft contribution is of order $m \alpha^5 \log(\alpha)$ and $m \alpha^5$.
Note that the ultrasoft contribution has been renormalized in the $\MS$ scheme and its $\mu$ dependence cancels
against that one of the matching coefficient $c_D$ [see Eq.~\eqref{cdcsd2}] in the NLO potential of Eq.~\eqref{E-NLO}.

Finally, $C^{\rm nad}_{\kappa\kappa'}(\r)$ is the {\em nonadiabatic coupling}~\cite{RevModPhys.35.473,MarxHutter}:
\bea
C^{\rm nad}_{\kappa\kappa'}(\r) &=& \int d^3 z\, \phi^*_{\kappa}(\r;\z) [ - \bnabla^2_r/M, \phi_{\kappa'}(\r;\z)] 
\nn \\
&=& \int d^3 z\, \phi^*_{\kappa}(\r;\z) \left(-\frac{\bnabla^2_r}{M} \phi_{\kappa'}(\r;\z)\right)
+ \frac{2}{M} \int d^3 z \, \phi^*_{\kappa}(\r;\z) \left(-i\bnabla_r \phi_{\kappa'}(\r;\z)\right)\left(-i\bnabla_r  \right)\,.
\label{nonadia}
\eea   
\end{widetext}
The first integral in the second line is the matrix element of the kinetic energy operator of the relative motion of the nuclei,
it is of order $m\alpha^2 \, m/M \sim m \alpha^{7/2}$, 
and the second integral involves the momentum of their relative motion,
it is of order $m\alpha^2 \, (m/M)^{3/4} \sim m \alpha^{25/8}$. 
When the $\phi_{\kappa}$'s are real and $\kappa=\kappa'$, the second integral vanishes.

We conclude by commenting on some general features of the BOEFT.
First, we would like to notice that there is no extra approximation by writing $S(t,\r,\z)$ as in Eq.~(\ref{exp-HHl}), 
since the eigenfunctions $\phi_{\kappa}(\r;\z)$ form a complete set and the $\Psi_{\kappa}(t,\r)$ play the role of time-dependent expansion coefficients. 
However, as it is well-known in treatments employing the Born--Oppenheimer approximation, this is useful in practice only 
when the dynamics of the heavy degrees of freedom (with mass $M$) is much slower than the dynamics of the light degrees of freedom (with mass $m$), 
a feature that permits to define an adiabatic dynamics for the heavy particles and to treat departure from adiabaticity using perturbation theory in the small parameter $m/M \ll 1$,
as we have done above.
Otherwise, when $M \simeq m$, the concept of adiabatic motion for one of the particles loses sense and an expansion like Eq.~(\ref{exp-HHl}) would be useless.
A way to see this is by noticing that mixing terms in the energy levels of the BOEFT would count like $m\alpha^2$,
a fact that would prevent the separation of the electron from the nuclei dynamics.

\begin{figure}[ht]
\begin{center}
  \includegraphics[height=5cm]{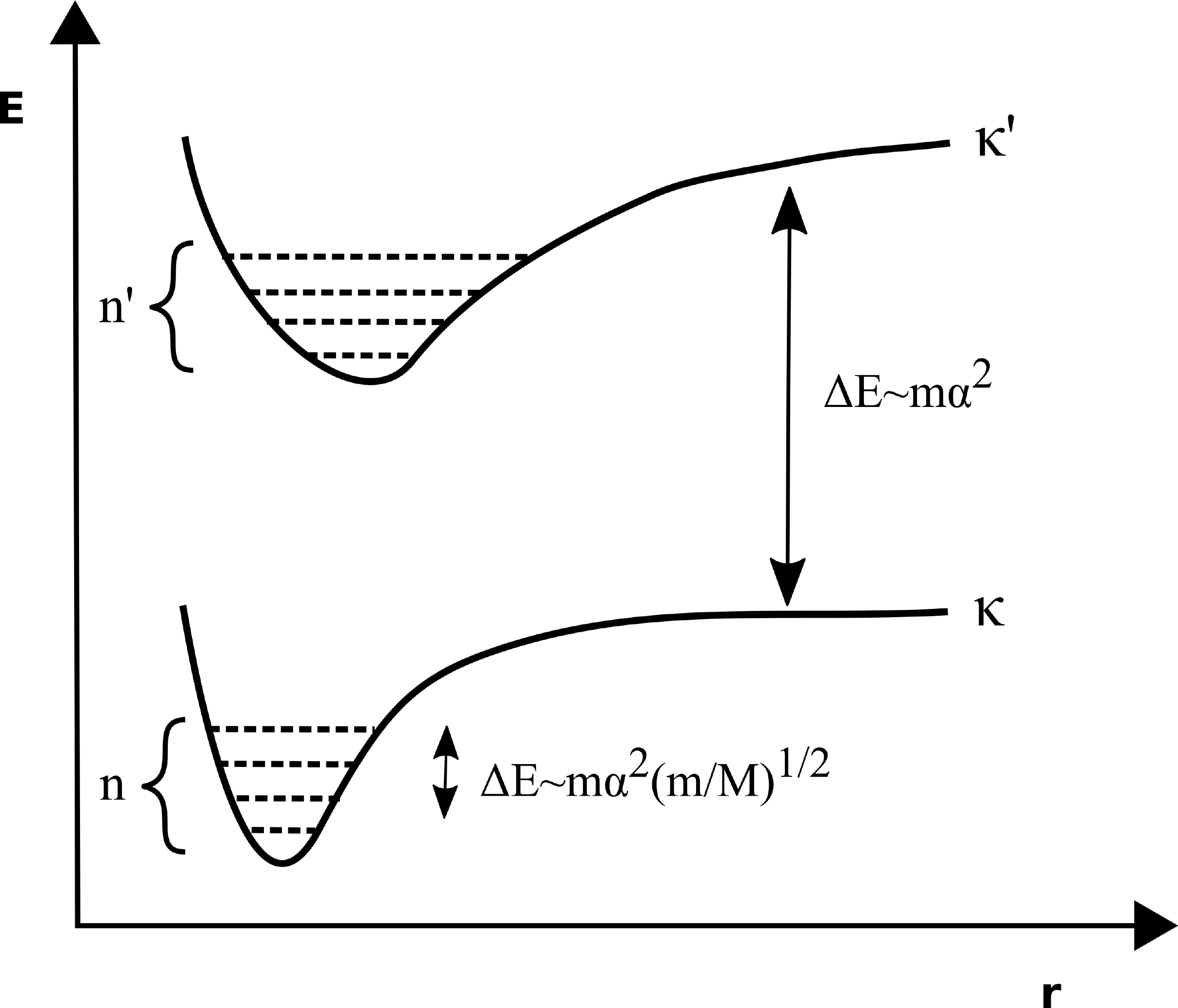}
  \caption{Sketch of the energy levels of a $H_2^+$-like molecule.\label{figlevels}}
\end{center}
\end{figure}

Under the adiabatic assumption the molecular energy levels are distributed as sketched in Fig.~\ref{figlevels}.
Electronic excitations define for each nuclei separation a potential $V^{\rm light}_\kappa(\r)$.
These potentials are separated by large gaps of order $m\alpha^2$.
For each electronic excitation, the nuclei motion induces smaller excitations of order $m\alpha^2\sqrt{m/M}$.
We can compute these smaller excitations in the BOEFT for each electronic potential $V^{\rm light}_\kappa(\r)$.
They are at leading order the eigenvalues of $H^{(0)}_\kappa$.
It is astounding that the wave functions of these nuclear vibrational modes can not only be computed but experimentally directly visualized: 
for the $H_2^+$ ground state potential $V^{\rm light}_0(\r)$ see~\cite{H2exp}.

\section{Power counting in the BOEFT}
\label{sec:self}
In this section we examine in detail the power counting of the BOEFT that we have just developed.
The main aim is to substantiate the starting assumption in the construction of the BOEFT, namely that the kinetic term $-\bnabla^2_r/M\ll m\alpha^2$.
Also of interest is the size of the nonadiabatic coupling.

The derivative $\bnabla_r$ can act on the nuclei fields $\Psi_{\kappa}(t,\r)$ as well as on the electronic wave functions $\phi_{\kappa}(\r;\z)$.
The size of the derivative turns out to be different for nuclei and electrons.
In the case of $\bnabla_r$ acting on $\phi_{\kappa}(\r;\z)$, it scales like $\sim m v$. 
Since the electron is bound to the nuclei through Coulomb interactions, we have that $v \sim \alpha$. 
In the case that the derivative acts on $\Psi_{\kappa}(t,\r)$, it scales like $\sim M w$, where $w$ is the relative velocity of the nuclei.
Therefore, our goal is to asses the size of $w$.

Since the system is bound, the nuclei will have a stable equilibrium arrangement and oscillate around an average separation $r_0$. 
Without the electron the two nuclei would not form a bound state, hence $r_0$ is an emergent scale, whose size needs to be determined. 
Let us consider the ground-state electron energy ($\kappa = 0$) and expand the total potential $V(\r) = V^{\rm LO}_{ZZ}(\r) + V^{\rm light}_0(\r)$
around the equilibrium position $r_0$ (we have adjusted the potential so that its minimum is zero):
\beq
V(\r) \approx \frac{1}{2} K_0 \, (\r-\r_0)^2, \hspace{0.5cm}{\rm where}\hspace{0.5cm} K_0 = \bnabla^2_{r_0}V(\r_0)\,.
\eeq
The Hamiltonian of the relative motion is that of a harmonic oscillator. 
The ground-state energy $E_0$ is given by
\beq
E_0 = 3 \sqrt{\frac{K_0}{2M}}\,.
\eeq
The equilibrium position $r_0$ of the nuclei is determined from
\beq
\bnabla_{r_0} V(\r_0) = \bnabla_{r_0} \left( \frac{\alpha Z^2}{r_0}+ V_0^{\rm light}(\r_0)\right) = 0\,.
\label{equilibrium}
\eeq 
Because $V_0^{\rm light}(\r_0)$ is the ground state energy of Eq.~\eqref{eigen-HHl}, it is of order $m\alpha^2$ (${\cal O}(Z^2) \sim 1$).
Hence Eq.~\eqref{equilibrium} implies
\beq
r_0 \sim \frac{1}{m \alpha}\,.
\eeq 
That is, the average size of the nuclei separation is of the same order as the electron-nucleus separation. 
Clearly, this is a particular feature of the Coulomb interaction between the nuclei; 
for a different $r$ dependence of the nucleus-nucleus interaction, $r_0$ may be not of the order of the Bohr radius.

From the above result it follows that
\beq
K_0 = \bnabla^2_{r_0}V(\r_0) \sim m^3 \alpha^4\,,
\eeq
and that the ground-state vibrational energy is
\beq
E_0 \sim \left(\frac{m^3\alpha^4}{M}\right)^{1/2} = m\alpha^2 \sqrt{\frac{m}{M}}\,.
\label{E0-vib}
\eeq
Transitions between low-lying vibrational states are also of order $m\alpha^2\sqrt{m/M}$. 
We note that the scaling behavior of $E_0$ implies a large cancellation between $V^{\rm LO}_{ZZ}(\r)$ and $V^{\rm light}_0(\r)$ near the equilibrium position, 
since each of these two potentials scales like $m\alpha^2$.

The virial theorem for the harmonic oscillator relates the expectation value of the kinetic energy with the total energy, 
\beq
2\langle \Psi_0|\frac{\bnabla^2_r}{M}|\Psi_0\rangle = E_0 \,,
\eeq
from where the size of the kinetic-energy operator acting on $\Psi$ follows
\beq
\frac{\bnabla^2_r}{M}\sim m\alpha^2 \sqrt{\frac{m}{M}} \,.
\label{kinpsi}
\eeq
Our initial assumption was that the kinetic energy associated with the relative motion of the nuclei is small compared to the ultrasoft scale, 
from there we integrated out the latter and matched pNRQED to the BOEFT. 
The above analysis shows that the energy scale associated with the relative motion of the nuclei is indeed largely suppressed by a factor $\sqrt{m/M}\sim \alpha^{3/4}\approx 0.025$ 
with respect to the ultrasoft scale, which justifies the initial assumption.

The size of $\bnabla_r$ acting on $\Psi$ and the relative velocity of the nuclei follows from \eqref{kinpsi}:
\bea
\bnabla_r & \sim & m\alpha \left(\frac{M}{m}\right)^{1/4}\,, 
\label{nrpsi}\\
        w & \sim & \alpha \left(\frac{m}{M}\right)^{3/4}\,.
\eea
A more detailed look reveals, however, that the counting of Eq.~(\ref{nrpsi}) applies only to the radial component of $\bnabla_r$. 
Indeed, in spherical coordinates we have $\bnabla_r=\left(\partial_r,\,\partial_{\theta}/r,\,\partial_{\phi}/(r\sin \theta)\right)$, 
and since the angles are dimensionless variables, the size of the last two components is determined by $r\sim r_0 \sim 1/(m\alpha)$. 
This implies also that the counting \eqref{kinpsi} is appropriate for the radial part of the kinetic energy, 
whereas $-2/(Mr) \, \partial/\partial r \sim m \alpha^2 (m/M)^{3/4}$ and the angular part $\L^2/(Mr^2)$ scales like $m\alpha^2(m/M)$. 

\begin{figure}[ht]
\makebox[0.5truecm]{\phantom b}\includegraphics[height=6cm]{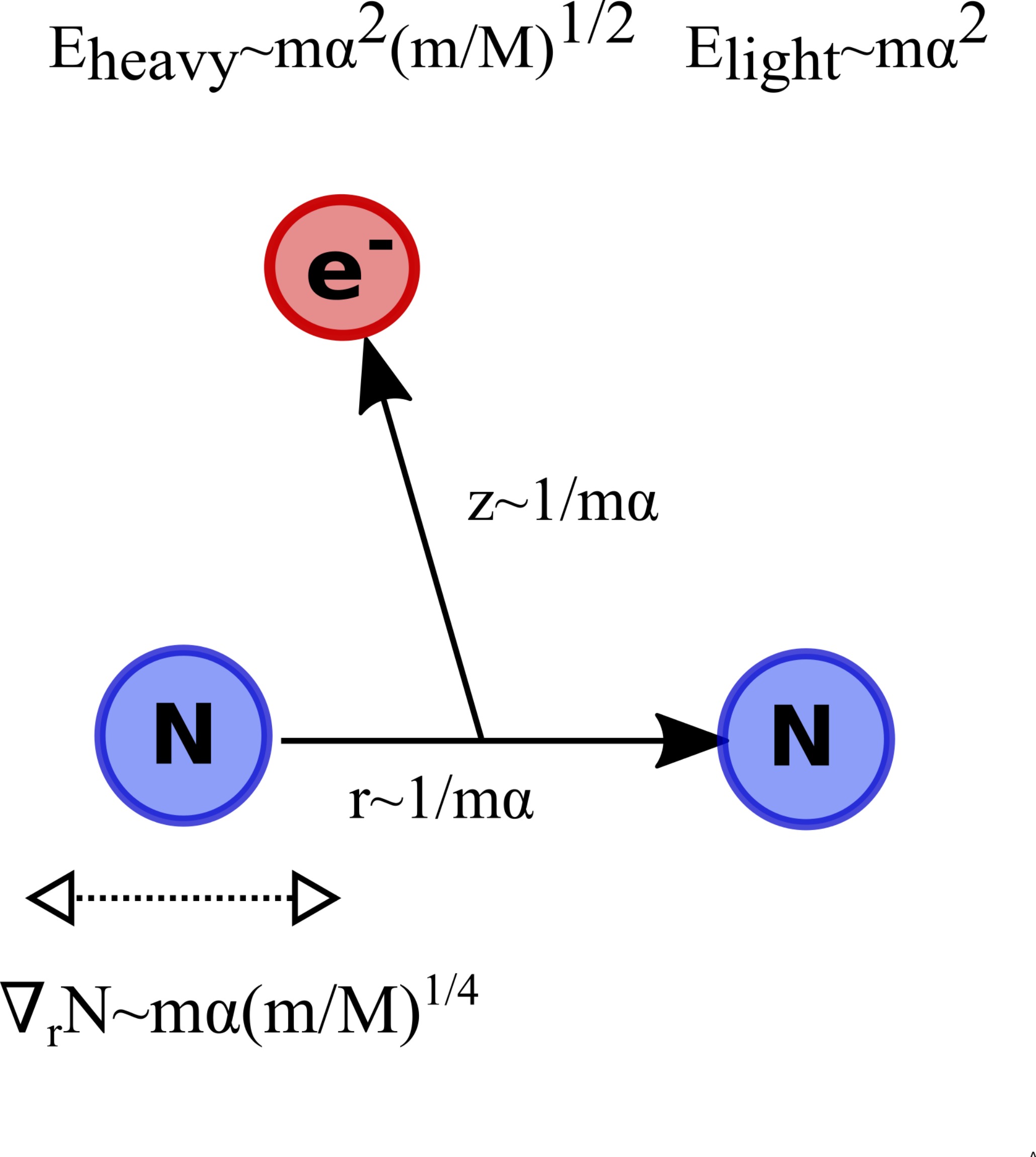}
  \caption{Energy scales for a $H_2^+$-like molecule.\label{figmolecule}}
\end{figure}

The size of the kinetic term in Eq.~(\ref{kinpsi}) sets the energy scale for the BOEFT.
Hence it determines the scaling of photon fields and derivatives acting on them. 
The last ingredient to complete the counting rules for the BOEFT is the scaling of $\bnabla_z\sim 1/\z\sim m\alpha$, which is inherited from pNRQED of Sec.~\ref{sec:pNRQED}.
The molecular energy scales are summarized in Fig.~\ref{figmolecule}.

We apply now the counting rules to the nonadiabatic coupling $C^{\rm nad}(\r)$ defined in \eqref{nonadia}. 
The largest contribution comes from the radial piece of the second term, which is of $\mathcal{O}\left(m\alpha^2(m/M)^{3/4}\right)$, 
while the first term and the angular piece of the second one are $\mathcal{O}\left(m\alpha^2(m/M)\right)$. 
Therefore, at leading order the nonadiabatic coupling can be neglected and the equation of motion for the field $\Psi_{\kappa}(t,\r)$ reads
\beq
i \partial_t \Psi_\kappa(t,\r) = H^{(0)}_\kappa(\r)\Psi_\kappa(t,\r)\,,
\label{boscheq}
\eeq
which is nothing else than the Schr\"odinger equation that describes the motion of the heavy particles in the {\em Born--Oppenheimer approximation}~\cite{bo1,bo2,bo3}. 
Equation~(\ref{boscheq}) produces the leading-order energy eigenvalues for the diatomic molecule, but it does not describe well the angular wave functions~\cite{RevModPhys.35.473}. 
This is a consequence of the angular piece of the kinetic term being of the same size as the angular parts of $C^{\rm nad}_{\kappa\kappa}$. 
The {\em adiabatic approximation}~\cite{RevModPhys.35.473,MarxHutter} corresponds to including in the above Schr\"odinger equation the diagonal term $C^{\rm nad}_{\kappa\kappa}(\r)$
\beq
i \partial_t \Psi_\kappa(t,\r)=\left[H^{(0)}_\kappa(\r)+C^{\rm nad}_{\kappa\kappa}(\r)\right]\Psi_\kappa(t,\r)\,.
\label{adbapp}
\eeq

One can use an iterative procedure to solve the problem: starting from the zeroth-order solution in which the nonadiabatic coupling $C^{\rm nad}$ is neglected, 
one can treat $C^{\rm nad}$ as a perturbation~\cite{2008JChPh.129c4102P} since its contribution to the energy is suppressed by an amount $(m/M)^{1/4} \approx 0.15$ with respect to the zeroth-order energy. 
We emphasize again that this relies on the Coulomb nature of the nucleus-nucleus interaction and on the smallness of the ratio $m/M$. 
Let $\Psi^{(0)}_{\kappa n}(\r) = \langle \r| \kappa n\rangle^{(0)}$ be the eigenfunctions of the leading-order Hamiltonian $H^{(0)}_\kappa(\r)$ with eigenvalues $E^{(0)}_{\kappa n}$,
then the corrections to $E^{(0)}_{\kappa n}$ can be computed in perturbation theory, so that
\beq
E_{\kappa n} = E^{(0)}_{\kappa n} + E^{(1)}_{\kappa n} + E^{(2)}_{\kappa n} + \dots\;.
\eeq
The leading-order correction $E^{(1)}_{\kappa n}$ comes from the diagonal nonadiabatic coupling and reads 
\beq
E^{(1)}_{\kappa n} =  {}^{(0)}\langle\kappa n|C^{\rm nad}_{\kappa\kappa}|\kappa n\rangle^{(0)}\,, 
\label{nadfo}
\eeq
where 
\beq
{}^{(0)}\langle \kappa n|O|\bar\kappa \bar n\rangle^{(0)} = \int d^3r \, \Psi^{(0)*}_{\kappa n}(\r)O(\r) \Psi^{(0)}_{\bar \kappa \bar n}(\r)\,.
\eeq
It is of order $m\alpha^2(m/M)^{3/4} \sim m \alpha^{25/8}$. The nondiagonal nonadiabatic coupling provides mixing with different electronic excitations.
The first contribution appears at order $m\alpha^2(m/M)^{3/2} \sim m \alpha^{17/4}$ and reads 
\beq
E^{\rm mix}_{\kappa n} = \sum_{\bar\kappa\neq k,\,\bar n}\frac{|{}^{(0)}\langle \kappa n|C^{\rm nad}_{\kappa\bar\kappa}|\bar\kappa \bar n\rangle^{(0)}|^2}{E_{\kappa n}-E_{\bar\kappa \bar n}} \,.
\label{nadso}
\eeq
More important than the mixing with states belonging to different electronic excitations is the mixing with states in the same one.
The mixing is in this case suppressed by a mere factor $(m/M)^{1/4}\sim \alpha^{3/8}$. 
We will not display here explicitly this kind of contributions that follow straightforwardly from time-independent quantum-mechanical perturbation theory.
We add that the recoil corrections to the electronic levels \eqref{E-rec} and \eqref{E-rec2} contribute first at order 
$m\alpha^2(m/M) \sim m \alpha^{7/2}$ and $m\alpha^2(m/M)^{2} \sim m \alpha^{5}$ respectively.
Finally, the NLO corrections to the electronic levels \eqref{E-NLO} contribute first at order $m\alpha^4$, 
while the ultrasoft corrections \eqref{Eultrasoft} contribute first at order $m\alpha^5 \log(\alpha)$ and $m\alpha^5$.

Let us now summarize the steps necessary for a numerical evaluation of the molecular energy levels using the BOEFT. 
First, the electronic static energies $V^{\rm light}_{\kappa}$ and wave functions $\phi_{\kappa}$ are obtained by solving the eigenvalue equation~\eqref{eigen-HHl} 
(see, for example, Ref.~\cite{Ponomarev:1967}). 
The BOEFT matching coefficients in Eqs.~\eqref{E-rec}-\eqref{Eultrasoft} and \eqref{nonadia} can then be evaluated. 
The nuclei wave functions $\Psi^{(0)}_{\kappa n}$ and eigenenergies $E^{(0)}_{\kappa n}$ are obtained by solving Eq.~\eqref{boscheq}, 
which requires the input of $V^{\rm light}_{\kappa}$ computed in the first step. 
The final values for the molecular energy levels are obtained by adding to $E^{(0)}_{\kappa n}$ in standard perturbation theory 
the corrections given by considering the higher-order operators in the Lagrangian of Eq.~\eqref{Lag-fin}.

\section{The BOEFT for QCD: heavy hybrids and adjoint tetraquark mesons} 
\label{sec:BO-QCD}
In the context of QCD, it exists a system analog to the QED diatomic molecule. 
It is the system formed by a heavy quark-antiquark pair and some light degrees of freedom that can be either gluonic or light quark in nature. 
Similarly to the QED bound state, the QCD system develops three well separated energy scales: 
the heavy-quark mass $M$ (hard scale), the relative momentum $M w$ (soft scale), where $w$ is the heavy-quark relative velocity, 
and the binding energy $M w^2$. 
Furthermore, there is the scale associated with nonperturbative physics, $\Lambda_{\rm QCD}$ that plays the role of the ultrasoft scale in the hadronic case. 
Restricting ourselves to the case $M w\gg\Lambda_{\rm QCD}$, we can use weakly-coupled pNRQCD~\cite{Pineda:1997bj,Brambilla:1999xf}
to describe the heavy quark-antiquark pair, which is called quarkonium if bound, pretty much in the same way as pNRQED, described in Sec.~\ref{sec:pNRQED}, 
can be used to describe electromagnetic bound states. However, a situation that has no analog in pNRQED, 
the heavy quark-antiquark fields can appear in pNRQCD either in a color-octet or in a color-singlet configuration.

At energies of the order of $\Lambda_{\rm QCD}$, the spectrum of QCD is formed by color-singlet hadronic states that are nonperturbative in nature. 
An interesting case it that one of exotic hadrons made of a color-octet heavy quark-antiquark pair bound with light degrees of freedom. 
Such a system can be studied similarly to the QED diatomic molecules. 
The heavy quarks play the role of the nuclei and the gluons and light quarks play the role of the electrons.

In a diatomic molecule the electrons are non-relativistic with energies of the order of the ultrasoft scale, $m \alpha^2$, 
whereas, as we have seen, the nuclei have a smaller energy due to their heavier mass. 
In a hadron made of a color-octet heavy quark-antiquark pair, the light degrees of freedom are relativistic with a typical energy and momentum of order $\Lambda_{\rm QCD}$.
This implies that the typical size of the hadron is of the order of $1/\Lambda_{\rm QCD}$.
If the mass of the heavy quarks is much larger than $\Lambda_{\rm QCD}$, there may be cases where 
also the typical momentum $M w$ of the heavy quarks in the hadron is larger than~$\Lambda_{\rm QCD}$.
The scaling of the typical distance of the heavy quark-antiquark pair depends on the details of the full inter-quark potential,
which has a long-range nonperturbative part and a short-range Coulomb interaction.
It may therefore happen that the heavy quark and antiquark are more closely bound than the light degrees of freedom.
This situation is interesting because the hadron would present a hierarchy between the distance of the quark-antiquark pair and the typical size of the light degrees of freedom 
that does not exist in the diatomic molecular case where the electron cloud and the two nuclei have the same size.
A consequence of this is that while the molecule is characterized by a cylindrical symmetry, 
the symmetry group of the hadron would be a much stronger spherical symmetry at leading order in a (multipole) expansion in the distance of the heavy quark-antiquark pair.
This modifies significantly the power counting of the hadronic BOEFT with respect to the molecular one leading to new effects.
In order to emphasize the difference between the hadronic and molecular case, we will assume in the following  
that the typical distance between the heavy quark and antiquark is of order $1/(Mw)$.

\begin{figure}[h]
\makebox[0.3truecm]{\phantom b}\includegraphics[height=5cm]{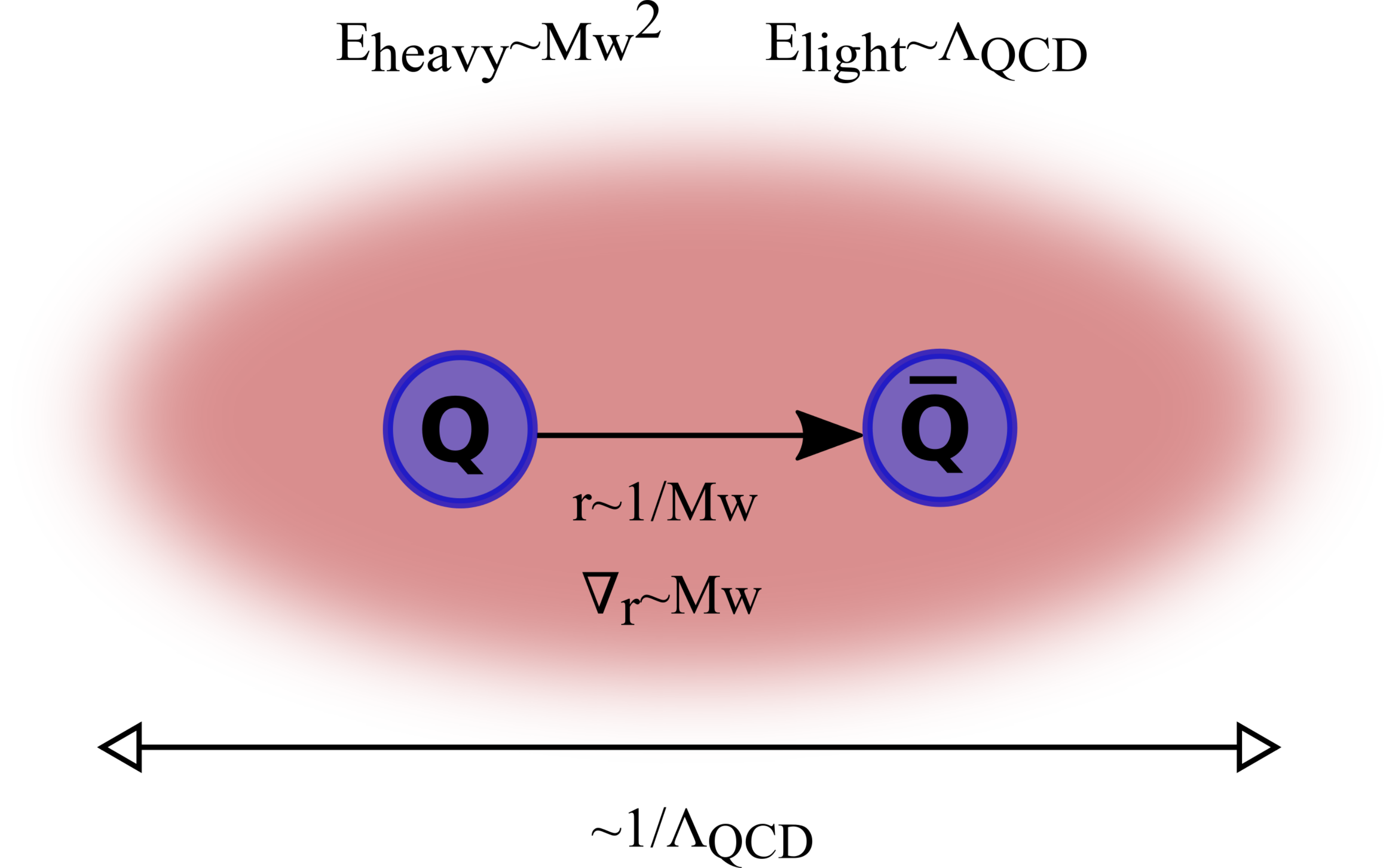}
  \caption{Energy scales of a quarkonium hybrid or tetraquark.\label{figexotic}}
\end{figure}

The kinetic energy associated with the relative motion of the quark-antiquark pair scales like $M w^2$.
If we look at hadrons that are in the ground state or in the first excited states only, we may require that $M w^2 \ll \Lambda_{\rm QCD}$.
As we have seen discussing the diatomic molecule, in order for a Born--Oppenheimer picture to emerge and for the BOEFT to provide a valuable theory  
it is crucial that the excitations between the heavy particles happen at an energy scale that is smaller than the energy scale of the light degrees of freedom.
In summary, we will require the following hierarchy of energy scales to hold true: $M w \gg \Lambda_{\rm QCD} \gg M w^2$~\cite{Brambilla:1999xf}. 
The different energy scales are shown in Fig.~\ref{figexotic}.

After integrating out the hard and soft scales from QCD and projecting on quarkonium states, 
one arrives at the pNRQCD Lagrangian in the weakly-coupled regime, which at leading order in $1/M$ and at $\mathcal{O}(r)$ 
in the multipole expansion is (we neglect the light-quark masses and higher-order radiative corrections to the dipole operators)
\begin{eqnarray}
&& L_{\rm pNRQCD} = \int d^3R\,\bigg\{ -\frac{1}{4} G_{\mu \nu}^a G^{\mu \nu\,a} + \sum^{n_f}_{i=1}\bar{q}_i i\slashed{D} q_i
\nn \\
&& + \int d^3r \,\Big( \text{Tr}\left[S^{\dag}\left(i\partial_0-h_s\right)S + O^{\dagger}\left(iD_0-h_o\right)O\right]  
\nn \\
&& \hspace{1.1cm}
+ g\text{Tr}\left[O^{\dagger}\bm{r}\cdot\bm{E}S+S^{\dagger}\bm{r}\cdot\bm{E}O\right]
\nn \\
&& \hspace{1.1cm}
+\frac{g}{2}\text{Tr}\left[O^{\dagger}\bm{r}\cdot\bm{E}O+O^{\dagger}O\bm{r}\cdot\bm{E}\right] \Big)\bigg\}\,,
\label{pnrqcd1}
\end{eqnarray}
where $S$ and $O$ are the heavy quark-antiquark color-singlet and color-octet fields respectively normalized with respect to color. 
They depend on $t$, $\bm{r}$, the relative coordinate, and $\bm{R}$, the c.m. position of the heavy quark-antiquark pair. 
All the fields of the light degrees of freedom in Eq.~(\ref{pnrqcd1}) are evaluated at $\bm{R}$ and~$t$; 
in particular, $G^{\mu \nu\,a} = G^{\mu\nu\,a}(\bm{R},\,t)$, $q_i = q_i(\bm{R},\,t)$ and $iD_0 O = i \partial_0O-g\left[A_0(\bm{R},\,t),O\right]$. 
The field $\bm{E}$ is the chromoelectric field, $G^{\mu \nu\,a}$ the gluonic field strength tensor and $q_i$ are light-quark fields appearing in $n_f$ flavors.
The singlet and octet Hamiltonians read (in the c.m. frame) 
\begin{align}
h_s=&-\frac{\bnabla^2_r}{M}+V_s(r)\,,
\\
h_o=&-\frac{\bnabla^2_r}{M}+V_o(r)\,,
\end{align}
where $V_s(r) = -4\als/(3r) +  \dots$ and $V_o(r) = \als/(6r) +  \dots$ are the color-singlet and color-octet potentials respectively; $\als$ is the strong coupling.

The Lagrangian \eqref{pnrqcd1} is the analog of the Lagrangian \eqref{LNpsigi-HHl} for diatomic molecules. 
The difference is that in the Lagrangian \eqref{pnrqcd1} the number of gluons and light quarks is not fixed as the number of electrons is in~\eqref{LNpsigi-HHl}. 
This stems from the fact that the electrons are nonrelativistic, which implies that their number is conserved at the low energy of pNRQED, 
while gluons and light quarks are massless relativistic particles and thus their creation and annihilation are still allowed in the Lagrangian~\eqref{pnrqcd1}.
 
The Hamiltonian density corresponding to the light degrees of freedom at leading order in $1/M$ and in the multipole expansion is
\begin{align}
h_{0}(\bm{R}) = \frac{1}{2} \left(\bm{E}^a\bm{E}^a+\bm{B}^a\bm{B}^a\right) - \sum^{n_f}_{f=1}\bar{q}_f \, i  \bm{D}\cdot{\bm \gamma} \, q_f \,.
\label{h-light}
\end{align} 
It plays the same role as the Hamiltonian density of Eq.~(\ref{h0}) does for the diatomic molecule. 
As anticipated, the symmetry groups of the two Hamiltonians are, nevertheless, different: 
the Hamiltonian density in Eq.~(\ref{h0}) has a cylindrical symmetry, while Eq.~(\ref{h-light}) has a spherical symmetry. 
The color-octet $G_{\kappa}^{ia}(\R)$ operators that generate the eigenstates of $h_0(\bm{R})$
\begin{align}
h_{0}(\bm{R})G^{ia}_{\kappa}(\bm{R})|{\rm US}\rangle = \Lambda_{\kappa} G^{ia}_{\kappa}(\bm{R})|{\rm US}\rangle\,,
\end{align}
form a basis of octet light degrees of freedom operators, labeled by the light-flavor $f$ and $J^{PC}$ quantum numbers, 
\begin{align}
\kappa = \{J^{PC},\,f\}\,,
\end{align}
and an extra label $i$ for states belonging to the same $J^{PC}$ representation. 
Note that the energy eigenvalue $\Lambda_{\kappa}$ is in general a complex number, whose imaginary part accounts for the possible decay of the state.

If we introduce the states ($O = \sqrt{2}O^aT^a$)
\begin{align}
|\kappa\rangle = O^{a\,\dagger}\left(\bm{r},\bm{R}\right) G_{\kappa}^{ia}(\bm{R})|{\rm US}\rangle\,,
\label{eigen1}
\end{align}
which are eigenstates of the octet sector of the pNRQCD Hamiltonian at leading order in the multipole expansion with eigenvalues $h_o+\Lambda_{\kappa}$, 
we can now project the Lagrangian of \eqref{pnrqcd1} onto the Fock subspace spanned by
\begin{align}
\int d^3r \, d^3 R \, \sum_{i\kappa}|\kappa\rangle \, \Psi^i_{\kappa}(t,\,\bm{r},\,\bm{R}) .
\label{boexpqcd}
\end{align}
This step is the equivalent for the hadronic system to the projection on the state of Eq.~\eqref{state12} and the expansion \eqref{exp-HHl} for the diatomic molecule.

Using Eq.~\eqref{boexpqcd} and integrating out light degrees of freedom of energy of order $\Lambda_{\rm QCD}$
we derive the BOEFT Lagrangian that describes the heavy quark-antiquark pair physics at the scale $M w^2$.
Since we are interested in bound states we will not consider sectors of the Lagrangian that describe transitions
between states with different $\kappa$ and decays into singlet states.
Up to next-to-leading order in the multipole expansion the Lagrangian reads
\begin{eqnarray}
L_{\rm BOEFT} &=& \int d^3R\, d^3r \, \sum_{\kappa} \, \Psi^{i\dagger}_{\kappa }(t,\,\bm{r},\,\bm{R})
\bigl[\left(i\partial_t - h_o - \Lambda_{\kappa}\right)\delta^{ij} \nn \\
&&-\sum_{\lambda}P^{i}_{\kappa\lambda}b_{\kappa\lambda}r^2P^{j\dagger}_{\kappa\lambda}+ \cdots \bigr] 
\Psi^j_{\kappa}(t,\,\bm{r},\,\bm{R})\,,
\label{bolag1}
\end{eqnarray}
where $P^i_{\kappa\lambda}$ are projection operators along the heavy-quark axis of the light degrees of freedom operator
(an implicit sum is understood over repeated $i$, $j$ indices).
There is one projection operator for each $-|j|\le\lambda\le|j|$.
These operators select different polarizations of the wave function $\Psi_{i\kappa}$.
For example, in the case of $J=1$ the operators are given by
\begin{align}
P^{l}_{10}&=\hat{r}^l \,,\label{pr10}\\
P^{l}_{1\pm1}&=\left(\hat{\theta}^l\pm i\hat{\phi}^l\right)/\sqrt{2}\,,\label{pr11}
\end{align}
with
\begin{align}
\hat{r}&=(\sin(\theta)\cos(\phi),\,\sin(\theta)\sin(\phi)\,,\cos(\theta))^T \,, \nonumber \\
\hat{\theta}&=(\cos(\theta)\cos(\phi),\,\cos(\theta)\sin(\phi)\,,-\sin(\theta))^T \,, \nonumber \\
\hat{\phi}&=(-\sin(\phi),\,\cos(\phi)\,,0)^T\,.
\end{align}
For higher $J$ the projection operators can be built by multiplying $|j|$ powers of \eqref{pr10} and \eqref{pr11} with appropriate symmetrization of the indices
(see also~\cite{tetra}).
The projection operators are necessary to organize the states in Eq.~\eqref{eigen1} according to the quantum numbers of the exotic hadron.
In particular they project the light degrees of freedom operator onto the heavy quark-antiquark axis.
The quantum numbers of the exotic hadron are the same as the ones of the diatomic molecule presented in Sec.~\ref{sec:HHL} plus charge conjugation:
as we discussed, at leading order in the multipole expansion the symmetry of the hadron is spherical, hence the projectors commute with the eigenstates
of $h_0$ (the equivalent statement is not true in the molecular case), but higher-order terms break this symmetry to the original cylindrical one.
In Eq.~\eqref{bolag1}, the next-to-leading order term in the multipole expansion is $P^{i}_{\kappa\lambda} b_{\kappa\lambda}r^2P^{j\dagger}_{\kappa\lambda}$, 
whereas the dots stand for higher-order terms.

\begin{figure}[ht]
\includegraphics[height=4cm]{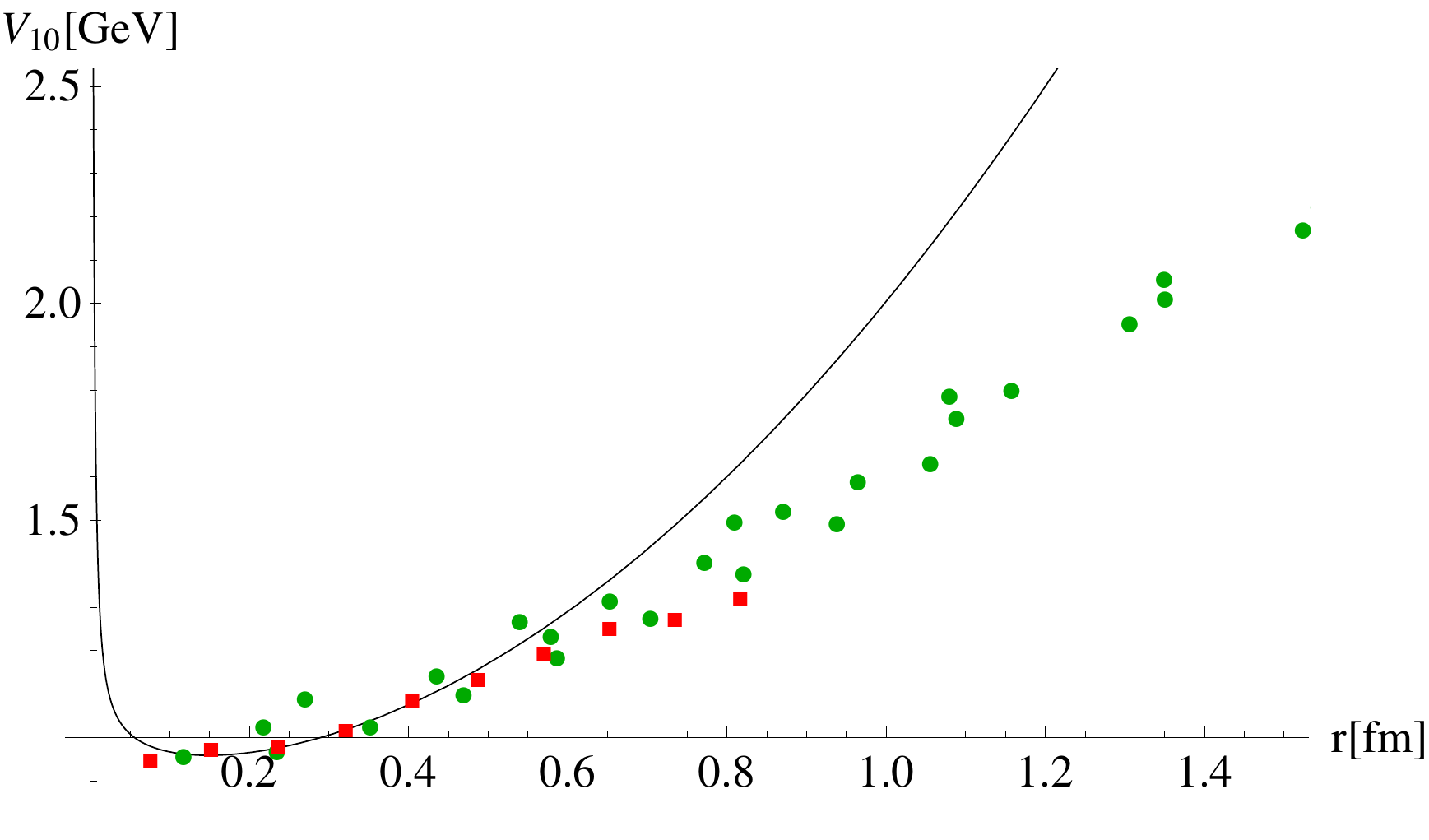}
\hspace{0.1cm}
\includegraphics[height=4cm]{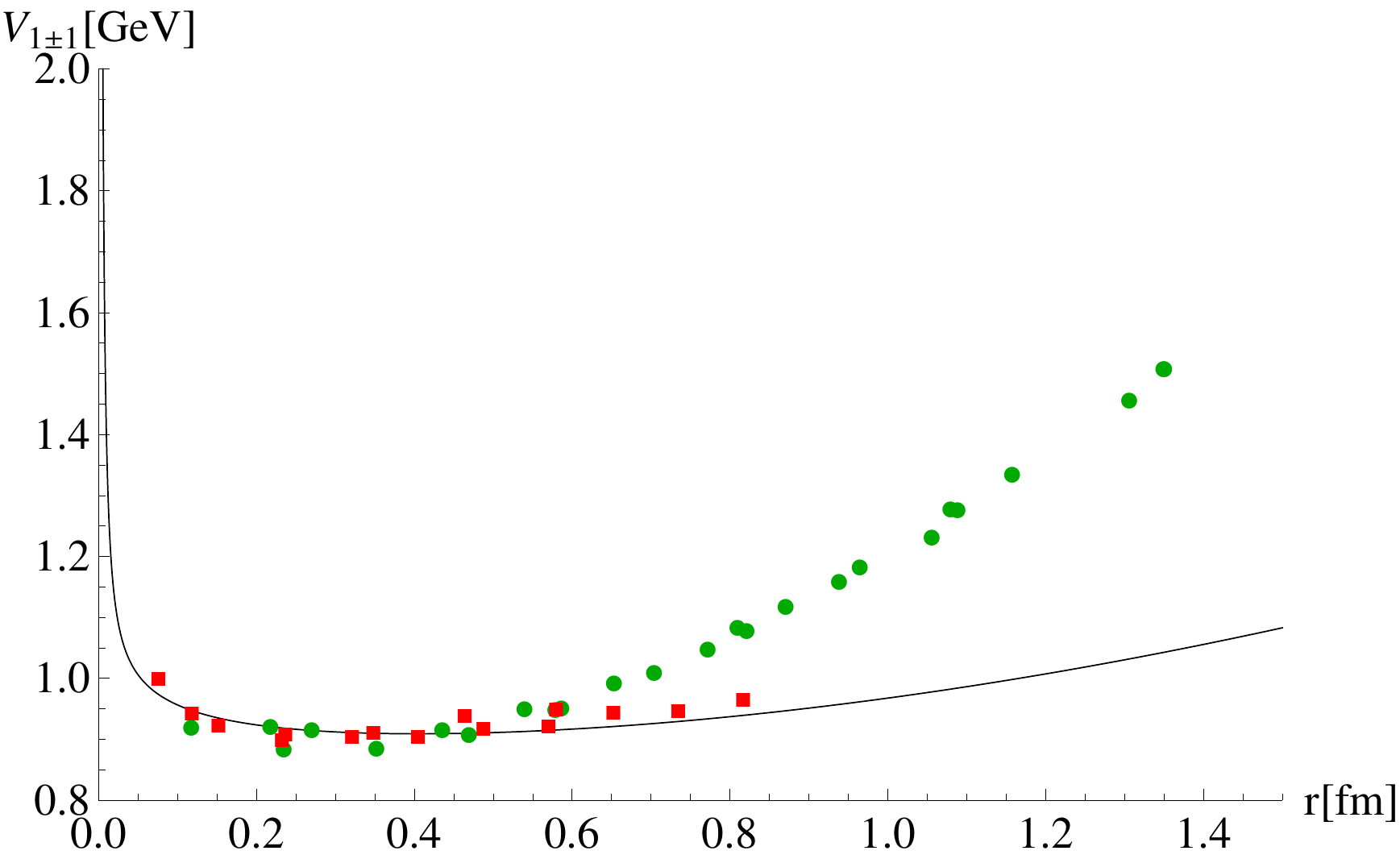}
\caption{Comparison of the hybrid quarkonium static energies generated by the lowest mass gluelump ($\kappa=1^{+-}$) 
computed on the lattice in Refs.~\cite{Bali:2003jq} (red squares) and \cite{Juge:2002br} (green dots) 
compared to the BOEFT static potential up to next-to-leading-order (solid black line), $V_{\kappa\lambda}=V_o(r)+\Lambda_{\kappa}+b_{\kappa\lambda}r^2$. 
The octet potential is taken in the Renormalon Subtracted (RS) scheme and up to $\als^3$. 
The mass of lowest laying gluelump is computed also in the RS scheme $\Lambda^{\rm RS}_{1^{+-}}=0.87$~GeV~\cite{Bali:2003jq}.  
The $b_{\kappa\lambda}$ coefficients are fitted to the lattice data for $r\lesssim 0.5$~fm yielding the values $b_{10}=1.112$~GeV/fm$^2$ and $b_{1\pm1}=0.110$~GeV/fm$^2$.
For lattice determinations of higher laying gluelump masses and static energies see Refs.~\cite{Griffiths:1983ah,Campbell:1984fe,Ford:1988ki,Perantonis:1990dy,Lacock:1996ny,Juge:1997nc,Juge:2002br,Bali:2000vr,Dudek:2007wv}.
\label{hypot}}
\end{figure}

The specific value of the next-to-leading-order term, $P^{i}_{\kappa\lambda} b_{\kappa\lambda}r^2P^{j\dagger}_{\kappa\lambda}$,
depends on nonperturbative physics and is unknown, however some of its characteristics can be determined on general grounds.
This term has its origin in the chromoelectric dipole interactions of Eq.~\eqref{pnrqcd1},
which couple the light degrees of freedom operator $G^{ia}_{\kappa}$ to the octet field giving corrections to the (static) energy of the system.
That this kind of corrections shows up for the static energy is a specific feature of QCD~\cite{Brambilla:1999qa,Brambilla:1999xf},
however, for nonstatic nuclei dipole interactions are also responsible for the Lamb shift of the diatomic molecule, as we have seen.
The $r^2$ dependence arises from the necessity of having at least two chromoelectric dipoles in order to conserve the $J^{PC}$ quantum numbers of~$G^{ia}_{\kappa}$.
Cylindrical symmetry and charge conjugation also imply $b_{\kappa\lambda}=b_{\kappa-\lambda}=b_{\kappa\Lambda}$. 
In Fig.~\ref{hypot} we show static potentials for the case of quarkonium hybrids, 
that is, for the case in which the considered light degrees of freedom are purely gluonic. 
The potentials correspond to $\kappa=1^{+-}$ and are compared to the static energies computed on the lattice in the quenched approximation. 
The values of $b_{\kappa\lambda}$ are fitted to the lattice data for $r\lesssim 0.5$fm.

Defining the projected wave function as 
\begin{align}
\Psi_{\kappa\lambda} =  P^{i\dagger}_{\kappa\lambda}\Psi^i_{\kappa}\,,
\end{align}
and using 
\begin{align}
\Psi^i_{\kappa}=\sum_{\lambda}P^{i}_{\kappa\lambda} \Psi_{\kappa\lambda}\,,
\end{align}
we can rewrite Eq.~\eqref{bolag1} as 
\bea
L_{\rm BOEFT} &=& \!\int d^3R\, d^3r \, \sum_{\kappa\lambda\lambda^{\prime}} 
\Psi^{\dagger}_{k\lambda}(t,\,\bm{r},\,\bm{R}) \biggl\{\Big[ i\partial_t - V_o(r) - \Lambda_{\kappa}
  \nn \\ 
  && \hspace{-1cm}
  -b_{\kappa\lambda} r^2 + \cdots \Big]\delta_{\lambda\lambda^{\prime}} + P^{i\dagger}_{\kappa\lambda}\frac{\bnabla^2_r}{M}P^{i}_{\kappa\lambda^{\prime}}\biggr\}
\Psi_{\kappa\lambda^{\prime}}(t,\,\bm{r},\,\bm{R}).
\label{bolag2}
\eea
The last term can be split into a kinetic operator acting on the heavy quark-antiquark field and a nonadiabatic coupling
\beq
P^{i\dagger}_{\kappa\lambda}\frac{\bnabla^2_r}{M}P^{i}_{\kappa\lambda^{\prime}}=\frac{\bnabla^2_r}{M} + C^{\rm nad}_{\kappa\lambda\lambda^{\prime}}\,,
\eeq
with 
\beq
C^{\rm nad}_{\kappa\lambda\lambda^{\prime}}=P^{i\dagger}_{\kappa\lambda}\left[\frac{\bnabla^2_r}{M}, P^i_{\kappa\lambda^{\prime}}\right]\,,
\eeq
being the nonadiabatic coupling analog to Eq.~\eqref{nonadia} for the diatomic molecule.

At this point it is important to review the sizes of the different terms appearing in Eq.~\eqref{bolag2}.
All dimensional quantities that arose from integrating out $\Lambda_{\rm QCD}$ are of order $\Lambda_{\rm QCD}$ to their dimension.
Hence $\Lambda_{\kappa}$ is of order $\Lambda_{\rm QCD}$ and $b_{\kappa\lambda}$ is of order $\Lambda_{\rm QCD}^3$.
The temporal derivative, the kinetic term and the potential up to the constant shift $\Lambda_{\kappa}$ are of order $M w^2$.
Unlike in the diatomic molecule case, $\bnabla_r$ has the same size for radial and angular pieces, 
because the momentum of the heavy quark is taken to scale like the inverse of the distance, $r$, between the quark and the antiquark. 
For the nonadiabatic coupling $C^{\rm nad}_{\kappa\lambda\lambda^{\prime}}$, the radial piece of the derivative $\bnabla_r$
acting on the projection operators $P^i_{\kappa\lambda^{\prime}}$ vanishes, since they do not depend on $|\r|$.
According to our counting, the size of the angular piece $\left[\L^2/(M r^2),\,P^i_{\kappa\lambda^{\prime}}\right]$ is $M w^2$, i.e.,
of the same order as the kinetic operator of the heavy quarks. This is different from the diatomic molecular case.

The equations of motion for the fields $\Psi_{\kappa\lambda}(t,\,\r,\,\bm{R})$ that follow from the Euler--Lagrange equation at leading order
are nothing else than a set of coupled Schr\"odinger equations
\begin{align}
i\partial_t \Psi_{\kappa\lambda}(t,\,\bm{r},\,\bm{R})=&\left[\left(-\frac{\bnabla^2_r}{M} + V_o(r) 
+ \Lambda_{\kappa}+b_{\kappa\lambda}r^2\right)\delta_{\lambda\lambda^{\prime}} \right.
\notag\\
&\left.-\sum_{\lambda^{\prime}}C^{\rm nad}_{\kappa\lambda\lambda^{\prime}}\right]\Psi_{\kappa\lambda^{\prime}}(t,\,\bm{r}, \,\bm{R}) \,.
\label{coupledhadron}
\end{align}
By solving them we obtain the eigenvalues $\mathcal{E}_N$ that give the masses $M_N$ of the states as 
\beq
M_N = 2M + \mathcal{E}_N\,.
\eeq

In summary, the spectrum of exotic hadrons that are sufficiently tightly bound that our hierarchy of scales,
and in particular the multipole expansion, applies is similar to that one of diatomic molecules illustrated in Fig.~\ref{figlevels}.
The quantum number $\kappa$ identifies, through different shifts $\Lambda_\kappa$, different excitations of the light degrees of freedom.
The gap between different excitations is (at least for the lower states) of order $\Lambda_{\rm QCD}$.
In the case of the diatomic molecule the different electronic excitations are separated by a gap of order $m\alpha^2$.
For each BO potential the vibrational modes of the heavy quark-antiquark pair generate a fine structure of levels, $\mathcal{E}_N$,
separated for fixed $\kappa$ by small gaps of order $Mw^2$. Similarly, in the molecular case the vibrational
modes of the nuclei induce small splittings of order $m\alpha^2 \sqrt{m/M}$.
There are, however, also noteworthy differences. In the hadronic case, if the size of the hadron is much larger
than the distance between the heavy quark and antiquark, then $\kappa$ labels spherically symmetric states.
Because the symmetry of the hadron is cylindrical, this means that at short distances 
some excitations of the light degrees of freedom turn out to be degenerate. As a consequence the equations of motion
are the coupled Schr\"odinger equations of Eq.~\eqref{coupledhadron} that mix different excitations, labeled by
$\lambda$, $\lambda'$, with the same $\kappa$. The mixing happens through the nonadiabatic coupling,
which under our assumptions counts like the quark-antiquark kinetic energy.
A physical consequence of the mixing is the so-called {\it $\Lambda$-doubling}, i.e., a lifting of degeneracy between
states with the same parity~\cite{Berwein:2015vca}.
In the molecular case, the size of the molecule and the typical distance between the nuclei is of the same order.
Because there is no special hierarchy between these two lengths there is
neither a special symmetry at short distance nor a corresponding degeneracy pattern. 
The equation of motion for the molecular case is the simple Schr\"odinger equation ~\eqref{boscheq} [or~\eqref{adbapp} in the adiabatic approximation].
In this case, different electronic excitations do not mix at leading order.
Moreover, the nonadiabatic coupling is subleading with respect to the relative kinetic energy of the nuclei.

\begin{figure*}[ht]
\centerline{\includegraphics[width=0.8\textwidth]{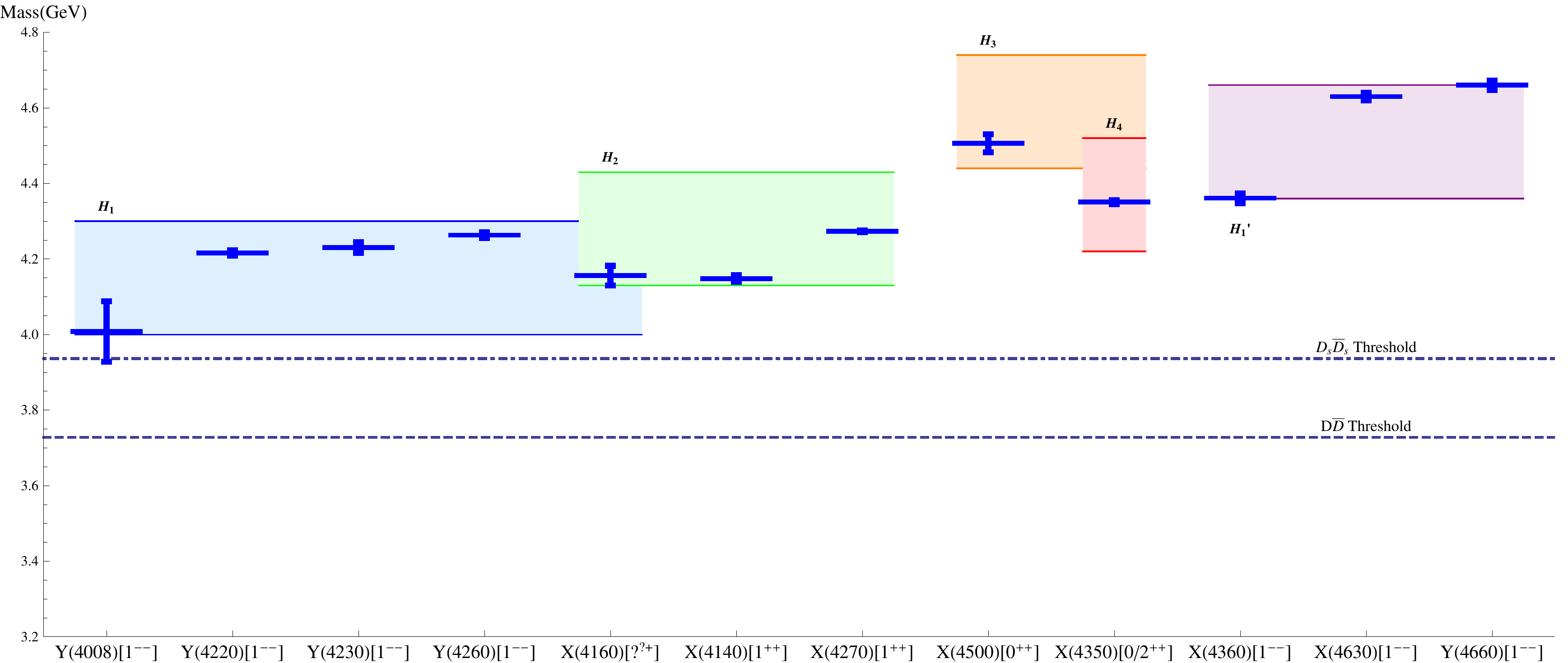}}
\caption{Comparison of the mass spectrum of neutral exotic charmonium states 
with the mass spectrum of charmonium hybrids computed in Ref.~\cite{Berwein:2015vca} using the BOEFT. 
The experimental states are plotted in solid blue lines with error bars corresponding to the average of the lower and upper mass uncertainties. 
Results of Ref.~\cite{Berwein:2015vca} are given in terms of spin-symmetry multiplets 
corresponding to solutions of the coupled Schr\"odinger equations~\eqref{coupledhadron} for different angular momentum, $l$, and parity. 
The spin-symmetry multiplets are labeled $H_1$ ($l=1$, positive parity), $H_2$ ($l=1$, negative parity), $H_3$ ($l=0$, positive parity), 
$H_4$ ($l=2$, positive parity) and $H^{\prime}_1$ (first radially excited state with $l=1$ and positive parity). 
The multiplets have been plotted with error bands corresponding to a gluelump mass uncertainty of $0.15$~GeV.\label{exp}}
\end{figure*}

The masses for heavy hybrid states have been obtained in Ref.~\cite{Berwein:2015vca} following the method just described.
There, the light-quark part of $h_0$ was omitted. 
In Fig.~\ref{exp} we reproduce the results of Ref.~\cite{Berwein:2015vca} compared with an updated list of possible experimental candidates.
Tetraquarks were discussed in Ref.~\cite{Braaten:2014qka} in the context of the BO approximation (see also~\cite{tetra}). 
In~\cite{Braaten:2014qka}, preliminary estimates for their masses were given assuming that the tetraquark static energies have the same shape as the hybrid ones
and using values for $\Lambda_\kappa$ from Ref.~\cite{Campbell:1985kp}.
One major difficulty is the lack of knowledge of the static energies carrying light-quark flavor quantum numbers.
One expects that lattice QCD will soon provide results on these and other crucial nonperturbative matrix elements to be used in the BOEFT developed here.

\section{Conclusions and perspectives} 
\label{sec:concl}
The Born--Oppenheimer approximation is the usual tool for solving the Schr\"odinger equation of molecules.
It relies on the movement of the nuclei being much slower than that of the electrons,
a circumstance that allows to study the electronic eigenstates and energy levels for fixed positions of the nuclei, the so-called static energies.
The wave functions of the molecule can then be expanded in terms of these electronic eigenfunctions
resulting in a Schr\"odinger equation describing the molecular energy levels.
We have used this hierarchy of scales to build an EFT that systematically describes the energy levels of the simplest diatomic molecule, $H^+_2$.

Our starting point has been an EFT of QED for the ultrasoft scale, pNRQED, adapted to the case of two nuclei and one electron.
Since pNRQED for two heavy and one light particle has not been presented in the literature before, we have worked out its derivation in some detail.
Particular care has been put in including all the relevant operators suppressed in powers of $m/M$, where $m$ and $M$ are the electron and nuclei masses respectively.
Counting $m/M\sim \alpha^{3/2}$ we have derived the pNRQED Lagrangian relevant to compute the spectrum up to $\mathcal{O}(m\alpha^5)$.

The assumption that the nuclei move slower than the electrons, which is at the basis of the Born--Oppenheimer approximation,
is equivalent to take the kinetic term of the nuclei to be of a smaller size than the energy scale of the electron dynamics, the ultrasoft scale.
Being these two scales well separated, it is natural in an EFT framework to integrate out the ultrasoft degrees of freedom
in order to obtain an EFT that describes the molecular degrees of freedom only.
We have carried out this integration obtaining a molecular EFT that we have named Born--Oppenheimer EFT (BOEFT).
Up to $\mathcal{O}\left(m\alpha^4\right)$ it is sufficient to match pNRQED and BOEFT at tree level, or equivalently,
to expand the matter field in the pNRQED Lagrangian in eigenfunctions of the leading-order Hamiltonian for the electron,
as it is done in the Born--Oppenheimer approximation of the Schr\"odinger equation.
Loop diagrams involving ultrasoft photons start contributing at $\mathcal{O}\left(m\alpha^5\right)$,
the first of such contributions being responsible for the $H^+_2$ molecular Lamb shift.
We have computed the leading ultrasoft loop and obtained the BOEFT Lagrangian relevant to compute the spectrum up to $\mathcal{O}\left(m\alpha^5\right)$.

The precise size of the nuclei kinetic operator has been obtained using the virial theorem to relate it to the potential acting on the nuclei.
At leading order this potential is formed by the repulsive Coulomb potential between the nuclei and the attractive electronic static energies.
Since the system is bound, the nuclei do not move over the whole size of the molecule, but oscillate around the minimum of the potential.
The size of the kinetic operator of the nuclei is of the order of $m\alpha^2\sqrt{m/M}$, which is smaller than the ultrasoft scale $m\alpha^2$.
This is consistent with the original statement that the two nuclei dynamics occurs at a lower energy scale than the electronic one.
The size of the nonadiabatic coupling could also be assessed resulting in the conclusion
that for diatomic molecules its contribution to the energy levels is suppressed by a factor $(m/M)^{1/4}$.

In the present paper we have derived the BOEFT Lagrangian for the $H^+_2$ molecule up to operators relevant for the spectrum up to  $\mathcal{O}\left(m\alpha^5\right)$.
This can be systematically improved by including higher-order operators in the power counting detailed in Sec.~\ref{sec:self},
and computing their corresponding matching coefficients.
Similarly, all the relevant contributions up to a certain precision to a specific observable can be determined with the help of the power counting,
which may be of crucial importance to handle high-precision calculations.

Having set the general framework for constructing the BOEFT in QED, we have analyzed systems in QCD analog to the diatomic molecule.
These are systems made of a heavy quark-antiquark pair, which plays the role of the heavy degrees of freedom,
bound with light-quarks or excited gluonic states, playing the role of the light degrees of freedom.
In particular, we have studied the case in which the quark-antiquark pair appears in a color-octet state.
In the short distance regime, $r\ll 1/\Lambda_{\rm QCD}$, the multipole expansion is applicable and the system can be described using weakly-coupled pNRQCD.

The energy scale of the leading-order light degrees of freedom dynamics is $\Lambda_{\rm QCD}$, while,
as in the molecular case, the heavy degrees of freedom dynamics, in this case that of the heavy quark-antiquark pair,
takes place at the lower energy scale $M w^2$.
We have identified the leading-order Hamiltonian in the multipole and $1/M$ expansions for the light degrees of freedom, $h_0$,
and defined a basis of color-octet light degrees of freedom operators, which, together with the heavy quark-antiquark octet field,
generate hadronic (color-singlet) eigenstates of the pNRQCD Hamiltonian.
The $\Lambda_{\rm QCD}$ scale has been integrated out and pNRQCD matched into a QCD version of the BOEFT.
At LO in the multipole expansion the matching can be done by just projecting the octet sector of the pNRQCD Lagrangian on the basis of eigenstates of $h_0$.
At NLO the matching requires a full nonperturbative computation, nevertheless, some constraints on the form of the NLO term can be obtained
from the multipole expansion itself and the cylindrical symmetry that the system possesses at finite separation between the heavy quarks.
As in the diatomic molecular case, a nonadiabatic coupling between the heavy quarks and the light degrees of freedom arises from the matching procedure,
however, unlike in the molecular case, this does not need to be suppressed with respect to the kinetic operator.
Furthermore, the nonadiabatic coupling mixes states that in the short distance limit have degenerate potentials,
therefore the mixing has to be taken into account when solving the set of Schr\"odinger equations
that result from the Euler--Lagrange equations of the BOEFT.
As a result the phenomenon known as $\Lambda$-doubling in molecular physics~\cite{LandauLifshitz} is more prominent in the QCD case~\cite{Berwein:2015vca}.

The BOEFT has been used to obtain the masses of the quarkonium hybrids in Ref.~\cite{Berwein:2015vca} (see also~\cite{Oncala:2017hop}).
Preliminary studies on quarkonium tetraquarks using a similar framework based on the BO approximation were carried out in Ref.~\cite{Braaten:2014qka}.
A further analysis is in preparation~\cite{tetra}.
The EFT presented here could be straightforwardly extended to describe any system made of two heavy quarks bound adiabatically with some light degrees of freedom.
An example are doubly heavy baryons, i.e., states with two heavy quarks and one light-quark.
Experimentally, doubly heavy baryons have been first observed at the LHCb~\cite{Aaij:2017ueg}.
For a study of this system in the framework of pNRQCD, we refer to~\cite{Brambilla:2005yk}.
Another example are pentaquark states made of two heavy quarks and three light-quarks.
Candidates have been observed at the LHCb~\cite{Aaij:2015tga}, but a pNRQCD based study of these systems is still to be done.

\section*{Acknowledgements}
We thank Matthias Berwein for comments on Sec.~\ref{sec:BO-QCD}. This work has been supported by the DFG and the NSFC through funds provided to the Sino-German CRC 110 ``Symmetries and the Emergence of Structure in QCD'' and by the DFG cluster of excellence ``Origin and Structure of the Universe''. 
A grant from the bilateral agreement between Bayerische Hochschulzentrum f\"ur Lateinamerika (BAYLAT) of the Bayerischen Staatsministeriums f\"ur Bildung und Kultus,
Wissenschaft und Kunst (StMBW) and Fun\-da\-\c{c}\~ao de Amparo \`a Pesquisa do Estado de S\~ao Paulo (FAPESP), contracts Nos. 914-20.1.3 (BAYLAT) and 2013/50841-1 (FAPESP) is acknowledged. 
G.K. acknowledges a grant from the Bavarian State Ministry of Education, Science and the Arts through the TUM International Center Visiting Program 2016.
The work of G.K. has been partially financed by Conselho Nacional de Desenvolvimento Cient\'{\i}fico e Tecnol\'ogico - CNPq, Grant No. 305894/2009-9, 
and Fun\-da\-\c{c}\~ao de Amparo \`a Pesquisa do Estado de S\~ao Paulo Grant No. 2013/01907-0.

\appendix

\section{The Lamb shift in the \texorpdfstring{$H^+_2$}{H2+} molecule} 
\label{app:matching}
In this appendix we derive Eq.~\eqref{Eultrasoft} following closely Ref.~\cite{Pineda:1997ie}.
When replacing $S(t,\r,\z)$ with the expansion \eqref{exp-HHl} in the Lagrangian \eqref{LNpsigi-HHl}, we obtain the nucleus-photon interaction terms
\bea
L_{\Psi-A} &=& \int d^3r \, \sum_{\kappa \kappa'} \Psi^\dag_{\kappa}(t,\r) \Bigl[  e_{\rm tot} A_0(t,\0)  \delta_{\kappa \kappa'}
  \nn \\
&& + \, e \, \E(t,\0)\cdot \langle \r,\kappa|\z|\r,\kappa'\rangle \Bigr] \Psi_{\kappa'}(t,\r) ,
\label{Psi-A}
\eea
where we have used that $e_{\rm eff} = e + {\cal O}(\alpha^2)$ and replaced $e_{\rm eff}$ with $e$; $\langle \r,\kappa|\z|\r,\kappa'\rangle$ is the matrix element
\beq
\langle \r,\kappa|\z|\r,\kappa'\rangle = \int d^3z \, \phi^*_{\kappa}(\r;\z) \,\z\,\phi_{\kappa}(\r;\z).
\eeq

The correction to the energy eigenvalues of the molecule coming from these terms can be obtained from the two-point correlation function  
\bea 
i\Pi_{\kappa}(t,\r,t',\r')&& =\langle {\rm US}|T[\Psi_{\kappa}(t,\r)\Psi^\dag_{\kappa}(t',\r')]|{\rm US}\rangle
\nn\\
=&& \int^{+\infty}_{-\infty} \frac{dE}{2\pi} \, e^{-iE(t-t')}\, i\Pi_{\kappa}(E,\r,\r') .
\label{def-Psi-2p}
\eea 
Second-order perturbation theory leads to
\begin{widetext}
\bea
i\Pi_{\kappa}(E,\r,\r') &=& i\,\Pi^{(0)}_{\kappa}(E,\r,\r')
\nn \\
&& \, + \int d^3\bar{r} \, i\Pi^{(0)}_{\kappa}(E,\r,\bar\r) \Biggl[- e^2 \sum_{\bar \kappa} \langle \bar\r,\kappa|z^i|\bar\r,\bar\kappa\rangle \, I^{ij}(E- H^{(0)}_{\bar\kappa}(\bar\r)) \,
\langle \bar\r,\bar\kappa|z^i|\bar\r,\kappa\rangle \Biggr] \, i\Pi^{(0)}_{\kappa}(E,\bar\r,\r')\,,
\label{Dys-eq}
\eea
\end{widetext}
where $\Pi^{(0)}_{\kappa}(E,\r,\r')$ is the zeroth-order two-point correlation function, corresponding to the zeroth-order Hamiltonian $H^{(0)}_\kappa(\r)$:
\beq
\Pi^{(0)}_{\kappa}(E,\r,\r') = \Pi^{(0)}_{\kappa}(E,\r) \, \delta^3(\r-\r')\,,
\label{Pi0-kap}
\eeq
with
\beq
\Pi^{(0)}_{\kappa}(E,\r) =  \frac{1}{E - H^{(0)}_\kappa(\r) + i \eta}\,,
\eeq
and $I^{ij}(E)$ is the loop integral: 
\beq
I^{ij}(E) = \int \frac{d^4k}{(2\pi)^4} {\frac{i}{k^2}}\,\k^2\,\left(\delta^{ij} - \frac{k^i k^j}{\k^2}\right) \,\frac{i}{E - k^0 + i\eta} .
\label{loop-int}
\eeq
Note that in dimensional regularization the one-loop contribution induced by the vertex with the $A_0$ field in \eqref{Psi-A} vanishes. 

Using Eq.~(\ref{Pi0-kap}) into Eq.~(\ref{Dys-eq}) and integrating over $\bar\r$, we obtain ($\Pi_{\kappa}(E,\r,\r') = \Pi_{\kappa}(E,\r) \, \delta^3(\r-\r')$)
\bea
i\Pi_{\kappa}(E,\r) &=& i\,\Pi^{(0)}_{\kappa}(E,\r)
\nn\\
&&\hspace{-1cm}
+  i\Pi^{(0)}_{\kappa}(E,\r) \left[-i\Sigma_\kappa(E,\r)\right] \, i\Pi^{(0)}_{\kappa}(E,\r)\,,
\eea
where
\bea
\Sigma_\kappa(E,\r) &=& - i e^2 \sum_{\bar \kappa} \langle \r,\kappa|z^i|\r,\bar\kappa\rangle \,
\nn\\
&& \times I^{ij}(E - H^{(0)}_{\bar\kappa}(\r)) \, \langle \r,\kappa|z^j|\r,\bar\kappa\rangle\,.
\eea
The energy shift $\delta^{\rm US} E_\kappa(\r)$ is the self-energy calculated at $H^{(0)}_\kappa(r)$. Therefore, we need to evaluate
\bea
\Sigma_\kappa(H^{(0)}_\kappa(\r),\r) &=& - i e^2 \sum_{\bar \kappa} \langle \r,\kappa|z^i|\r,\bar\kappa\rangle\,
\nn\\
&& \hspace{-2cm}
\times I^{ij}(V^{\rm light}_\kappa(\r) - V^{\rm light}_{\bar\kappa}(\r))\,  \langle \r,\kappa|z^j|\r,\bar\kappa\rangle\,.
\eea

The loop integral $I_{ij}(E)$ in Eq.~(\ref{loop-int}) is ultraviolet divergent.
In dimensional regularization, it is given by (using the convention $D = 4 - \epsilon$)
\bea
I^{ij}(E) &=& - i \, E^3 \,\frac{\delta^{ij}}{6\pi^2} \, \left[\frac{1}{\epsilon} + \frac{1}{2} \log (4\pi) - \frac{\gamma_E}{2} \right.
\nn\\
&& \hspace{0.7cm}
\left.+ \log\left(\frac{\mu}{-E-i\eta}\right) + \frac{5}{6} - \log (2)\right].
\eea
The divergent part of the self-energy is then
\begin{widetext}
\bea
&&\Sigma^{\rm div}_{\kappa}(H^{(0)}_\kappa(\r),\r)  = - e^2 \frac{\delta^{ij}}{6\pi^2}\frac{1}{\epsilon}\,\sum_{\bar\kappa} \langle \r, \kappa |z^i|\r, \bar\kappa\rangle \,
(V^{\rm light}_{\kappa}(\r) - V^{\rm light}_{\bar\kappa}(\r))^3 \,\langle \r,  \bar\kappa |z^j|\r,\kappa\rangle 
\nn \\
&&\,\,= - e^2 \frac{\delta^{ij}}{6\pi^2}\,\frac{1}{\epsilon}\,\sum_{\bar\kappa} (V^{\rm light}_{\kappa}(\r) - V^{\rm light}_{\bar\kappa}(\r))\,\langle \r, \kappa|z^i (V^{\rm light}_{\kappa}(\r) 
- h_0(\r;\z))|\r,\bar\kappa \rangle\,\langle \r, \bar\kappa|(V^{\rm light}_{\kappa}(\r)- h_0(\r;\z)) z^j|\r,\kappa \rangle\,.
\eea
\end{widetext}
Now, since
\bea
&&\langle\r,\kappa|z^i (V^{\rm light}_{\kappa}(\r) - h_0(\r;\z))|\r,\bar\kappa \rangle 
\nn \\
&&= \langle \r,\kappa|[h_0(\r,\z),z^i]|\r,\bar\kappa\rangle  = - i \langle \r,\kappa|v^i_z|\r,\bar\kappa\rangle\,, 
\eea
where $v^i_z = - i \, \nabla^i_z/m$.
Then, $\Sigma^{\rm div}_{\kappa}(H^{(0)}_\kappa(\r),\r)$ can be written as
\bea
&&\Sigma^{\rm div}_{\kappa}(H^{(0)}_\kappa(\r),\r)= - e^2\frac{\delta^{ij}}{6\pi^2} \, \frac{1}{\epsilon} \sum_{\bar\kappa} \langle \r, \kappa |v^i_z|\r,\bar\kappa\rangle 
\nn\\
&& \hspace{2.5cm}
\times(V^{\rm light}_{\kappa}(\r) - V^{\rm light}_{\bar\kappa}(\r))\,\langle \r,\bar\kappa|v^j_z|\r,\kappa\rangle  
\nn\\
&& = - e^2  \frac{1}{6\pi^2} \frac{1}{\epsilon} \,\frac{1}{2}\langle \r,\kappa |[v^i_z,[ v^i_z,h_0(\r,\z)]]|\r,\kappa \rangle\,.
\eea
The double commutator acts only on the heavy-light Coulomb potentials $V^{LO}_{Ze}(\z\pm \r/2)$ in $h_0(\r,\z)$ {\textemdash} and the result is 
\bea
&&[v^i_z,[ v^i_z,h_0]]= -\frac{1}{m^2} [\nabla^i_z,[ \nabla^i_z, h_0]] 
\nn \\
&&\,\,= - \frac{Ze^2}{m^2} \, \left[\delta^3(\z-\r/2) + \delta^3(\z + \r/2)\right]\,.
\eea
Therefore, the divergence can be absorbed by the renormalization of the contact interaction term $V^{\rm ct}_{Ze}(\x)$ of Eq.~(\ref{VCT}).

Since the matching coefficients of the pNRQED Lagrangian were obtained in the $\MS$ renormalization scheme, we have to use the same scheme here; 
this amounts at subtracting the term $1/\epsilon + 1/2 \log (4\pi)-{\gamma_E}/{2}$ from $I^{ij}(E)$.
We are then left with
\bea
&&\Sigma_{\kappa}(H^{(0)}_\kappa(\r),\r) =
\nn\\
&& \quad 
- \frac{e^2}{6 \pi^2} \Biggl\{ - \frac{Z e^2}{2m^2}\,\left[ \log \left(\frac{\mu}{m}\right) + \frac{5}{6} - \log (2)\right] \rho_\kappa(\r)
\nn \\
&& \quad 
+ \sum_{\bar\kappa} |\langle \r,\kappa|{\bm v}_z|\r,\bar\kappa\rangle|^2\, (V^{\rm light}_{\kappa}(\r) - V^{\rm light}_{\bar\kappa}(\r)) 
\nn \\
&& \hspace{1cm}
\times\log\left(\frac{m}{-V^{\rm light}_{\kappa}(\r) + V^{\rm light}_{\bar\kappa}(\r)-i\eta}\right) \biggr\}\,,
\label{deltaE-fin}
\eea
where $\rho_\kappa(\r)$ is the electron density at the positions of the nuclei:
\beq
\rho_\kappa(\r) = |\phi_\kappa(\r;\z=\r/2)|^2 +|\phi_\kappa(\r;\z=-\r/2)|^2\,.
\eeq

The $\mu$ dependence in the second line of \eqref{deltaE-fin} cancels against the $\mu$ dependence of the matching coefficient $c_D$ in Eq.~(\ref{VCT}). 
Note that the last term is zero when $V^{\rm light}_{\bar\kappa}(\r) = V^{\rm light}_{\kappa}(\r)$. 
Also, there will be an imaginary part when $V^{\rm light}_{\kappa}(\r) > V^{\rm light}_{\bar\kappa}(\r)$,
indicating that the level $\kappa$ may decay to the level $\bar \kappa$.
The energy shift of the electronic states is given by the real part of $\Sigma_{\kappa}(H^{(0)}_\kappa(\r),\r) $
\bea
\delta E^{\rm US}_{\kappa}(\r) &=&  - \frac{e^2}{6 \pi^2} \biggl\{ - \frac{Z e^2}{2m^2}\, \left[ \log \left(\frac{\mu}{m}\right)+\frac{5}{6}-\log(2)\right] \, \rho_\kappa(\r) 
\nn \\
&& + \sum_{\bar\kappa} |\langle \r, \kappa |{\bm v}_z|\r,\bar\kappa\rangle|^2 \, (V^{\rm light}_{\kappa}(\r) - V^{\rm light}_{\bar\kappa}(\r)) 
\nn \\
&&\hspace{1cm}
\times \log\left(\frac{m}{|V^{\rm light}_{\kappa}(\r) - V^{\rm light}_{\bar\kappa}(\r)|}\right) \biggr\}\,.
\eea
This is precisely Eq.~(16) of Ref.~\cite{Pineda:1997ie}, if one identifies $V^{\rm light}_{\kappa}(\r)$ with the leading-order energies of the hydrogen atom, $E_n$,
and $\rho_e(\r)$ with the electron density~$|\phi_n(\0)|^2$.

\bibliographystyle{apsrev4-1}
\bibliography{BObib}

\end{document}